\documentclass[fleqn,usenatbib]{mnras}

\usepackage{newtxtext,newtxmath}

\usepackage[T1]{fontenc}
\usepackage[bottom]{footmisc}

\DeclareRobustCommand{\VAN}[3]{#2}
\let\VANthebibliography\thebibliography
\def\thebibliography{\DeclareRobustCommand{\VAN}[3]{##3}\VANthebibliography}

\usepackage{graphicx}	
\usepackage{amsmath}	
\usepackage{deluxetable}
\usepackage{pdflscape}

\title[L1521E COMs]{Detection of Complex Organic Molecules in Young Starless Core L1521E}

\author[S. Scibelli et al.]{
Samantha Scibelli,$^{1}$\thanks{E-mail: sscibelli@email.arizona.edu}
Yancy Shirley,$^{1}$
Anton Vasyunin$^{2}$
and Ralf Launhardt$^{3}$
\\
$^{1}$Steward Observatory, University of Arizona.
Tucson, AZ 85721 \\
$^{2}$Ural Federal University, Ekaterinburg, Russia \\
$^{3}$Max-Planck-Institut f\"{u}r Astronomie K\"{o}nigstuhl   17, D9117 Heidelberg, Germany \\
}

\date{Accepted XXX. Received YYY; in original form ZZZ}

\pubyear{2020}

\begin{document}
\label{firstpage}
\pagerange{\pageref{firstpage}--\pageref{lastpage}}
\maketitle

\begin{abstract} 
Determining the level of chemical complexity within dense starless and gravitationally bound prestellar cores is crucial for constructing chemical models, which subsequently constrain the initial chemical conditions of star formation. We have searched for complex organic molecules (COMs) in the young starless core L1521E, and report the first clear detection of dimethyl ether (CH$_3$OCH$_3$), methyl formate (HCOOCH$_3$), and vinyl cyanide (CH$_2$CHCN). Eight transitions of acetaldehyde (CH$_3$CHO) were also detected, five of which (A states) were used to determine an excitation temperature to then calculate column densities for the other oxygen-bearing COMs. If source size was not taken into account (i.e., if filling fraction was assumed to be one), column density was underestimated, and thus we stress the need for higher resolution mapping data. We calculated L1521E COM abundances and compared them to other stages of low-mass star formation, also finding similarities to other starless/prestellar cores, suggesting related chemical evolution. The scenario that assumes formation of COMs in gas-phase reactions between precursors formed on grains and then ejected to the cold gas via reactive desorption was tested and was unable to reproduce observed COM abundances, with the exception of CH$_3$CHO. These results suggest that COMs observed in cold gas are formed not by gas-phase reactions alone, but also through surface reactions on interstellar grains.
Our observations present a new, unique challenge for existing theoretical astrochemical models.
\end{abstract}




\section{Introduction} \label{sec:intro}

The chemistry of star and planet formation begins with the synthesis of molecules in interstellar molecular clouds. Dense ($\sim$ 10$^5$\,cm$^{-3}$) and cold (10\,K) regions within these clouds, called starless and gravitationally bound prestellar cores, represent the earliest observable stage of low-mass ($\mathrm{M} \leq \rm{few}$\,M$_\odot$) star-formation \citep{2007prpl.conf...17D, 2007ARA&A..45..339B, 2014prpl.conf...27A}. It is during this phase prior to the formation of a first hydrostatic core (the foremost stage of a protostar), that the initial conditions for star and planet formation are set. 

In the warm regions of early star formation, such as in hot cores and hot corinos, rich chemistry has been explained by a warm-up, gas-phase chemistry driven by the heat from the forming protostar that sublimates ice mantles
(\cite{2020ARA&A..58..727J} and references therein). Surprisingly, even in cold and shielded environments, rotational transitions of complex organic molecules (COMs) have been detected in the gas phase toward dense prestellar cores from deep millimeter wave observations \citep{2010ApJ...716..825O, 2012A&A...541L..12B, 2016ApJ...830L...6J, 2020ApJ...891...73S}. These large ($\geq6$\,atom) carbon-bearing molecules are of astrobiological significance, and their presence prior to protostar and planet formation provides insight into COM formation histories. 

There is still, however, a significant gap in our understanding of the development of COMs during this initial phase of low-mass star formation. The traditional picture depicts COMs forming in CO and H$_2$O rich ices that are irradiated by UV light and
cosmic rays \citep{2002ApJ...571L.173W, 2017MNRAS.467.2552C}. The subsequent formation of radicals in the warming ice, during the development of a hot corino, leads to the formation of complex organic species that are released into the gas phase when the young, accreting protostar heats up the dust grains above the sublimation temperature of the ices \citep{2006A&A...457..927G, 2008ApJ...674..984A, 2009ARA&A..47..427H}. Since prestellar cores have no internal heating source, several different scenarios of cold COMs formation have been proposed, including the desorption of radicals, yet their relative importance is not currently clear.

If COMs form in the ices, then the subsequent desorption of COMs in the outer, icy layers
of prestellar cores appears to occur at a rate fast enough to build up an observable 
abundance of COMs \citep{2014ApJ...795L...2V}.  Alternatively, COMs could form in the gas phase through neutral-neutral reactions from the chemical desportion of radicals formed in the ice \citep{2013ApJ...769...34V, 2017ApJ...842...33V}. New non-diffusive chemical models produce excited radicals instead by surface or bulk reactions, which leads to an enhancement of COM abundance in cold core conditions \citep{2020ApJS..249...26J}. In order to discern between different formation scenarios, we must understand the physical conditions in which COMs form as well as constrain their abundances.

The discovery of COMs in prestellar cores leads to fundamental questions of how quickly these COMs form, and to what level of complexity. In our previous study of a large sample ($31$) of starless and prestellar cores within the Taurus Molecular Cloud \citep{2020ApJ...891...73S}, we found a prevalence of the `mother molecule,' methanol (CH$_3$OH; $100$\% detection rate), which is the gas precursor of several COMs, as well as acetaldehyde (CH$_3$CHO; $70$\% detection rate).
For the larger molecules, like dimethyl ether (CH$_3$OCH$_3$) and methyl formate (HCOOCH$_3$), surveys have been limited to a few well-studied, evolved and dense objects, i.e., L1689B in Ophiuchus \citep{2012A&A...541L..12B} and L1544 in Taurus \citep{2016ApJ...830L...6J}. L1544 is the densest known prestellar core, with well developed collapse motions and a central density $> 10^{7} \mathrm{cm}^{-3}$ \citep{2019ApJ...874...89C}. Species such as CH$_3$O, CH$_3$CHO and CH$_3$OCH$_3$ are enhanced in L1544 by factors of $\sim2-10$ toward the CH$_3$OH peak with respect to (w.r.t.) the core's center \citep{2016ApJ...830L...6J}. It has been suggested that HCOOCH$_3$ and CH$_3$OCH$_3$ are synthesised in gas-phase reactions with CH$_3$OH and other precursor species ejected from grains via the process of reactive desorption \citep{2017ApJ...842...33V}. We know that CH$_3$OH and CH$_3$CHO are prevalent in the starless and prestellar cores in Taurus \citep{2020ApJ...891...73S}, now we look for more complex species.

In this study we targeted a dynamically and chemically young starless core in Taurus, Lynds 1521E (L1521E).
It has a modest central density of $2-3 \times 10^5$\,cm$^{-3}$, no kinematic evidence for collapse, little evidence for freezout of CO ($\lesssim$ factor of 2), and little evidence for chemically evolved indicators, such as abundant deuterated molecules \citep{2004A&A...414L..53T, 2011ApJ...728..144F}. L1521E can only have existed at its present density of $> 10^5 \mathrm{cm}^{-3}$ for less than $10^5$\,years, otherwise there would be more significant CO freezout (e.g., see \cite{2002MNRAS.337L..17R}). Uniquely, L1521E is also one of only two cores known in Taurus, the other Seo 32 in the B213 region, where the `early-time' chemical tracers (such as CCS and HC$_5$N) peak strongly at the same position as the dust continuum peak \citep{1992ApJ...392..551S, 2019ApJ...871..134S}.
Furthermore, ``late-time" chemical tracers such as NH$_3$, N$_2$H$^+$, and deuterated species 
are very weak \citep{2001ApJ...547..814H, 2019A&A...630A.136N}.  The chemical and kinematical evidence agree that this object is one of the youngest starless cores known, with densities at $> 10^5 \mathrm{cm}^{-3}$.  Thus, L1521E is an excellent object toward which to test the limits of how quickly COMs form, to test what level of chemical complexity COMs form, and to test at what molecular abundances COMs form.

We present detections and detailed column density analyses for multiple COMs towards the young starless core L1521E. In section \,\ref{sec:obs} we discuss the line observations made by the Arizona Radio Observatory (ARO) 12m telescope. In section \,\ref{sec:data} we discuss our process for estimating the H$_2$ column density as well as the line analysis for each molecule detected. Included is a detailed description of non-detected transitions as well as discussion of glycoaldehyde (cis-CH$_2$OHCHO), which was targeted but not detected and known as an important molecule for building up sugars of astrobiological importance (see \cite{2020AsBio..20.1048J}). We also calculate column densities and account for the beam filling factor. We then discuss how our results compare to other sites of low-mass star formation, as well as to modeled abundances, in section \,\ref{sec:discussion}. Lastly, we summarize our results in section \,\ref{sec:conclusion}.

\section{ARO 12m Observations}\label{sec:obs}

Molecular line observations were made with the ARO 12m telescope during three separate seasons, two years apart, using two different backend receivers. The first observing shifts between January 12, 2017 and April 27, 2017 with 10 tunings between 84 and 102 GHz ($3.6 - 2.9 \mathrm{mm}$).  The sideband separating, dual polarization 3mm ALMA prototype receiver was used with the MAC backend in a mode with 300 MHz bandwdth with 48.8\,kHz resolution (bandwidth of $880 - 1060$\,km\,s$^{-1}$ with $0.14 - 0.17$\,km\,s$^{-1}$ resolution).  Two IFs within the same sideband were connected to the Vertical (Vpol) and Horizontal 
(Hpol) polarization feeds.  The lower sideband was used for frequencies less than $97$ GHz while the upper sideband was used for frequencies higher than $97$ GHz.  The rejection between sidebands was measured by injecting a tone into the opposite sideband and measuring the strength of the signal in the observed
sideband. The sideband rejection was $\geq 18$ dB for all tunings.  

The second set of observing shifts took place between May 12th, 2019 and June 25th, 2019 with 3 different tunings on the 3mm receiver between 84 and 99\,GHz ($3.6 -3.0 \mathrm{mm}$), and 4 tunings on the 4mm receiver between 69 and 77\,GHz ($3.8 - 4.3 \mathrm{mm}$). These observations used the new AROWS backend in a mode with 125\,MHz bandwidth and 19.5\,kHz resolution. There was a third set of observations that took place between February 17th and February 27th, 2019 utilizing the AROWS backend, tuned to the 79\,GHz HCOOCH$_3$ transitions with the 4mm receiver.

Position-switched observations of L1521E were centered on the $850$ $\mu$m dust continuum peak position of 4$^h$ 29$^m$ 14.9$^s$ $+$26$^d$ 13$^{\prime}$ 56.6$^{\prime\prime}$ J2000.0, with 5 to 6 minutes total integration time (ON+OFF) between calibration scans (longer in good weather). During the first observing run, measurements of Venus and Mars were used to determine the average beam efficiency. Values were consistent for all 3mm tunings with an average value of $\eta_\mathrm{mb, MAC, 3mm} = 0.87 \pm 0.01$. Additionally, a factor of 1.14 needed to be multiplied to the final main beam temperature due to systematic calibration error present in the software of the MAC. The MAC antenna temperature was found to be 14$\%$ lower than the new more trusted AROWS backend (see \cite{2020ApJ...891...73S}). For the two remaining observing runs, AROWS was used and average beam efficiencies for the 3mm and 4mm observations were calculated separately.
We note that during the Summer between the first observing run (MAC) and the second and third sessions (AROWS), the antenna surface accuracy was improved.
During the second run, we used observations of just Mars to get an average value of $\eta_\mathrm{mb, AROWS, 3mm} = 0.91\pm 0.01$ for all 3mm tunings. For the 4mm tunings, we average Venus, Mars and Jupiter measurements to determine an average beam efficiency of $\eta_\mathrm{mb, AROWS, 3mm} = 0.93\pm 0.01$.
In the third observing session, exclusively set to a 4mm tuning, the average beam efficiency was calculated to be $\eta_\mathrm{mb, AROWS, 4mm} = 0.92\pm 0.01$, using Jupiter and Mars measurements.

Each tuning was chosen to include one or more COMs within the bandpass. Simple LTE, or local thermal equilibrium, excitation analyses were performed to estimate the most likely line to be observed in the 3mm(4mm)
bandpass. In LTE the same excitation temperature is assumed for all transitions, which we have to assume is true because no collisional rates are available. In \cite{2020ApJ...891...73S} excellent agreement was found when both the LTE and non-LTE methods were applied to methanol, CH$_3$OH, which does have published collisional rates. Data reduction was performed using the CLASS program of the GILDAS package\,\footnote{\url{http://iram.fr/IRAMFR/GILDAS/}}. The AROWS spectra have been hanning smoothed by two channels since the default software writes out the data at half the actual resolution.

\begin{figure}
\centering
\begin{center}$
\begin{array}{c}
\includegraphics[width=80mm]{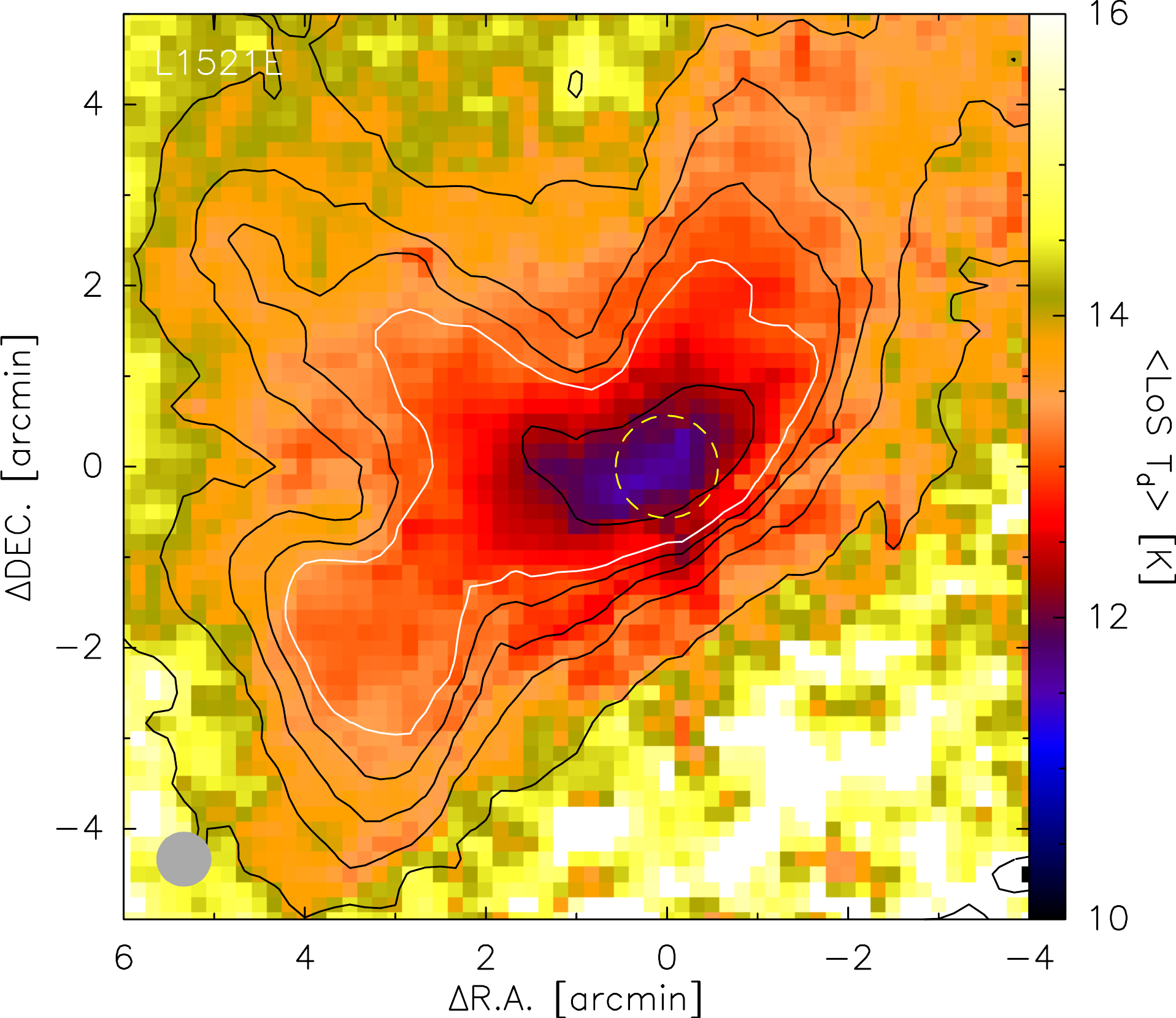} 
\end{array}$
\end{center}
\caption{\label{hersh} A map of the average line-of-sight dust temperature (color scale) and column density (contours) determined from SED fitting of \textit{Herschel Space Observatory} observations at $250, 350, 500$ $\mu$m centered on L1521E at 4$^h$ 29$^m$ 14.9$^s$ $+$26$^d$ 13$^{\prime}$ 56.6$^{\prime\prime}$ J2000.0.  The $\mathrm{N_{H}}$ contours correspond to $2, 4, 6, 8, 10, 20 \times 10^{21}$ cm$^{-2}$ (for $\mathrm{N_{H_2}}$, divide by 2). The yellow dashed circle indicates the average 12m beam size within which we find an average $\langle \mathrm{N_{H_2}} \rangle = 1.6 \times 10^{22}$ cm$^{-2}$. }
\end{figure}

\section{Analysis and Results} \label{sec:data} 

\subsection{Dust Continuum}

\cite{2013A&A...551A..98L} developed a continuum pixel-by-pixel SED
fitting technique to model the average line-of-sight dust temperature distribution and the H$_2$
column density (assuming a total gas mass to dust mass ratio of 150:1 and OH5 dust opacities,
\cite{1994A&A...291..943O}) from \textit{Herschel Space Observatory} and other FIR and submillimeter observations. 
L1521E was observed with MIPS on the \textit{Spitzer Space Telescope} at $160$ $\mu$m and with 
SPIRE on the \textit{Herschel Space Observatory} at $250, 350,$ and $500$ $\mu$m.  Unfortunately, no 
PACS observations of L1521E exist. MIPS observations have severe instrumental artifacts that
prohibit detailed, pixel-by-pixel modeling of that image but a total flux may be used as
a constraint when modeling the integrated SED \citep{2007ApJ...668.1042K}. L1521E was also observed with SCUBA at the 15m JCMT \citep{2005MNRAS.360.1506K}, but this map was too small to recover the emission level from the extended envelope and could therefore not easily be combined with the \textit{Herschel} data. The data reduction, including color corrections, of the Herschel/SPIRE data is described in \cite{2013A&A...551A..98L}. The results of SED modeling the archival \textit{Herschel} data are shown in Figure\,\ref{hersh}.  

The central 
regions of L1521E have
peak column densities of $\mathrm{N_H} = 3.9 \times 10^{22}$\,cm$^{-2}$ and average line-of-sight dust
temperatures of $11$\,K. The average H$_2$ density within the typical 12m beam size 
of $1^{\prime}$ (corresponding to $\sim 8400$\,AU at 140\,pc) is $\mathrm{N_{H_2}} = 1.6 \times 10^{22}$\,cm$^{-2}$.
We used this average H$_2$ column density for all subsequent calculations of
molecular abundances. The corresponding mass is $2.2$\,M$_\odot$. Our SPIRE image modeling results are consistent within a factor of two in column density with the analysis of SPIRE-FTS observations by 
\cite{2016MNRAS.458.2150M}.

Using the $850$\,$\mu$m JCMT map and the same techniques and grid of radiative transfer models described in \cite{2005ApJ...632..982S}, we find that a thermally-supported Bonnor-Ebert sphere with a central density of $3 \times 10^5\,\mathrm{cm}^{-3}$ is the best-fitted model. This result is consistent with our archival \textit{Herschel} SED modeling and the published radiative transfer modeling of MAMBO $1.2$ mm data from  \cite{2004A&A...414L..53T} ($n(r) = 2.7 \times 10^5\, \rm{cm}^{-3} / [1 + (r/6.3 \times 10^{16}\, \rm{cm})^2]$. 
We use these density profiles to calculate the virial parameter, $\alpha_{vir} = 3R\sigma_v^2/a_1GM(R)$.
The line-of-sight velocity dispersion, $\sigma_v = 0.25 \pm 0.02$ km s$^{-1}$, is calculated from the average and standard deviation of the observed linewidths toward the dust continuum peak in \cite{2019A&A...630A.136N} and is corrected for the molecule specific thermal broadening and the sound speed
\citep{1992ApJ...384..523F}. The density profile correction factor, $a_1$, is calculated using the techniques described in the appendices of \cite{1992ApJ...395..140B} and \cite{2011A&A...535A..49S}.
The exact radius of L1521E is uncertain and depends on how the core boundary is defined (i.e. for examples of various definitions see \citealt{1989ApJS...71...89B, 2015MNRAS.450.1094P, 2015ApJ...805..185S, 2019ApJ...877...93C}).
Assuming that it is in between angular radii of $60^{\prime\prime}$ to $140^{\prime\prime}$, then we find that $\alpha_{vir}$ is between $1.8 - 0.9$ which would indicate that L1521E is near the typical dividing line between a starless core and a gravitationally-bound prestellar core.
We shall continue to use the more general term starless when referring to the L1521E core.

\subsection{Molecular Identifications}

With confidence, we report the first detection of CH$_3$OCH$_3$, HCOOCH$_3$ and CH$_2$CHCN in the young starless core L1521E. Additionally, eight transitions of CH$_3$CHO and seven transitions of CH$_2$CHCN were detected, each set used to construct rotation diagrams to constrain excitation temperature and column density. The properties of the observed lines are listed in Table\,\ref{COMresults}. For each transition, we calculated the integrated intensity over the line and its uncertainty from the following equation, 
\begin{equation}
I \pm \sigma_I = \sum_i T_{i} \delta v_{ch} \pm \sigma_T \sqrt{\frac{3 \Delta v_{\rm{FWHM}}  \delta v_{ch}}{\sqrt{2 \ln{2}}}} \;\;,
\end{equation}
where $\delta v_{ch}$ is the spectrometer velocity resolution, and $\sigma_{T}$ is the baseline rms. The integrated intensity and uncertainty are calculated over a velocity range corresponding to
the $\pm 3\sigma$ width of a Gaussian with the same integrated intensity. This technique works well
for lines that are Gaussian and do not have significant skewness (for instance due to outflows), which is the case for all lines observed toward L1521E. The median FWHM for our lines is at $0.35$\,km\,s$^{-1}$ (see Table \,\ref{COM_FWHM}). We note that in Table \,\ref{COM_FWHM} there is slight discrepancy in v$_\mathrm{LSR}$ for the two transitions of CH$_2$CHCN which were detected with the old (less trustworthy) MAC backend receiver. Otherwise, the v$_\mathrm{LSR}$ of the remaining lines lie at $\sim$ 7\,km s$^{-1}$ (median value of 6.97\,km s$^{-1}$). In the following subsections we present an overview of the molecules detected.

\begingroup
\onecolumn 

\begin{deluxetable}{clllllcccc}
\small
\tablecaption{Complex Organic Molecule Fit Results\label{COMresults} }
\tablewidth{0pt}
\tablehead{
\colhead{Molecule} & \colhead{Transition} & \colhead{$\nu$} & \colhead{E$_u$/k} & \colhead{$^{a}$g$_u$ } & \colhead{A$_{ul}$} & \colhead{T$_\mathrm{mb}$} & \colhead{$\sigma$(T$_\mathrm{mb}$)} & \colhead{I(T$_\mathrm{mb}$)} & \colhead{$\sigma$(I)} \\ 
\colhead{} & \colhead{ } & \colhead{(GHz) } & \colhead{(K)} & \colhead{} & \colhead{(s$^{-1}$)} & \colhead{ (mK)} & \colhead{(mK)} & \colhead{(mK km s$^{-1}$)} & \colhead{(mK km s$^{-1}$)}}
\startdata
CH$_3$CHO & $3_{1,3}-2_{0,2}$ A$^{*}$ & 101.89241 & 	7.7	& 14 &	4.0E-06	& 23.0 & 4.0	&	10.7	&	1.3\\
  & $5_{0,5}-4_{0,4}$ A & 95.96347 & 13.8   & 22& 3.0E-05 & 89.0 & 9.0  & 30.0 &   2.0 \\
  & $5_{0,5}-4_{0,4}$ E & 95.94744 & 13.9  &22 & 3.0E-05 & 51.0 & 8.0 & 23.0 & 2.0 \\
  & $2_{1,2}-1_{0,1}$ A & 84.21975 & 5.0  & 10 & 2.4E-06  & 23.0 & 4.0 & 7.1  & 0.8 \\
  & $4_{0,4}-3_{0,3}$ A & 76.87895 & 9.2  & 18& 1.5E-05 & 89.0 & 11.0 & 36.0 & 3.0 \\
 & $4_{0,4}-3_{0,3}$ E & 76.86644 & 9.3  & 18& 1.5E-05  & 103.0 & 10.0 & 41.0 & 3.0 \\
 & $4_{1,4}-3_{1,3}$ A & 74.89168 & 11.3  & 18& 1.3E-05& 53.0 & 11.0 & 14.0 & 2.0 \\ 
 & $4_{1,4}-3_{1,3}$ E & 74.92413 & 11.3  & 18& 1.3E-05& 48.0 & 9.0 & 13.0 &  2.0 \\
CH$_3$OCH$_3$ & $4_{1,4}-3_{0,3}$ AA & 99.32607 & 10.2  & 90 & 5.5E-06 &  9.1 & 3.0 & 2.9 &  0.6  \\
    & $4_{1,4}-3_{0,3}$ EE & 99.32522 & 10.2  & 144 & 5.5E-06 & 10.0 & 3.0 &  4.2 & 0.7 \\
    & $4_{1,4}-3_{0,3}$ AE+EA & 99.32436 & 10.2  & 90 & 5.5E-06 & 4.6 & 3.0 &  5.5 & 1.2\\
&	$4_{2,3} - 4_{1,4}$ EE $^{*}$	&	93.85710	& 	14.7	& 144	 & 3.6E-06	&	\nodata	& 		2.0	&	\nodata	&	0.8		\\
	&	$2_{2,1} - 2_{1,2}$ EE $^{*}$	&	89.69981	& 	8.4	& 
	 80	&  2.3E-06	&	\nodata	& 		2.0	&	\nodata	&	0.8		\\
HCOOCH$_3$	&	$8_{1,8} - 7_{1,7}$ A $^{*}$	&	89.31664	& 	20.1	&  34&	1.0E-05	&	\nodata & 4.0 & \nodata &  1.3	\\ 
&	$8_{1,8} - 7_{1,7}$ E $^{*}$	&	89.31466	& 	20.2& 34	& 	1.0E-05	&	\nodata	& 		4.0	& \nodata	&	 1.3	\\
    &	$7_{0,7} - 6_{0,6}$ A	& 79.78389	&  15.7 & 30	& 7.2E-06	& 11.0	&  3.0	& 6.4	&	 0.6 \\
	&	$7_{0,7} - 6_{0,6}$ E	& 79.78171	& 15.7 & 30	& 7.2E-06	&	12.0	& 	3.0		& 3.9	&	 0.9	\\
CH$_2$CHCN &	$11_{0,11} - 10_{0,10}$&	103.57539	& 29.9  &69	&   8.8E-05	&	\nodata  	&   3.0	& \nodata & 0.6  \\
&	$10_{1,9} - 9_{1,8}$&	96.98245	& 27.8  &63	&  7.2E-05	&	 14.0	&  3.0	& 4.8 & 0.6  \\
&	$10_{0,10} - 9_{0,9}$&	94.27664	& 24.9  &63	&  6.2E-05	&	 27.0	& 4.0 	& 8.0 & 0.9	\\
&	$10_{1,10} - 9_{1,9}$&	92.42626	& 	26.6  &	63&  6.8E-05	&	18.0	&  3.0	& 4.4 & 0.6\\
 &	$5_{1,5} - 4_{0,4}$ $^{*}$	&	89.13091	& 	8.8	  & 33 & 	1.8E-06	&	\nodata	& 	2.0	&	\nodata	&	0.8	\\
&	$9_{1,8} - 8_{1,7}$	$^{*}$&	87.31281	& 	23.1  &	57 & 	5.3E-05	&	24.0	& 5.0	&	9.9	&	1.9	\\
 &	$9_{0,9} - 8_{0,8}$ $^{*}$	&	84.94600	& 	20.4  &57	& 	4.9E-05	&	30.0	& 		4.0	&	12.0	&	1.7	\\
 & $8_{1,7}-7_{1,6}$ & 77.63384 & 18.9  &51 & 3.6E-05 & 37.0 & 5.0 & 12.5 & 1.3\\
 & $8_{0,8}-7_{0,7}$ & 75.58569 & 16.3  & 51 &  3.4E-05& 50.0 & 5.0 &  16.9  & 1.3 \\
 cis-CH$_2$OHCHO	&	$5_{2,4} - 4_{1,3}$ $^{*}$	&	87.76453	& 	10.7 & 11	& 	8.6E-06	&	\nodata	& 		4.0	&	\nodata	&	1.4		\\
	&	$8_{1,8} - 7_{0,7}$	$^{*}$&	85.78224	& 	18.9	& 17 &	1.5E-05	&	\nodata	& 		3.0	&	\nodata	&	1.2	\\
	&	$8_{0,8} - 7_{1,7}$	$^{*}$&	82.47067	& 	18.8	& 17 &	 1.3E-05	&	\nodata	& 4.0	 	&	\nodata	&	 1.3	\\
\enddata
\tablecomments{ $^{*}$Transitions targeted with the MAC backend. $^{a}$Values for CH$_3$CHO and HCOOCH$_3$ from JPL catalog\footnote{\url{https://spec.jpl.nasa.gov/}} \citep{1998JQSRT..60..883P} and for the remaining transitions from CDMS database\footnote{\url{https://cdms.astro.uni-koeln.de}} (\cite{2001A&A...370L..49M}, \cite{2005JMoSt.742..215M}, \cite{2016JMoSp.327...95E}). We note that A$_{ul}$ can be calculated by the following equation, A$_{ul}$ = (64$\pi^4 \nu_i^3$/3hc$^3$) (S$\mu^2$/(2J$'$ + 1)), where J$'$ is the upper J in the transition and line strength, S, can be interpolated from values tabulated in \cite{standards_microwave_1969}.}
\end{deluxetable}

\endgroup
\twocolumn

\subsubsection{Acetaldehyde (CH$_3$CHO)}

\begin{figure}
\centering
\begin{center}$
\begin{array}{c}
\includegraphics[width=80mm]{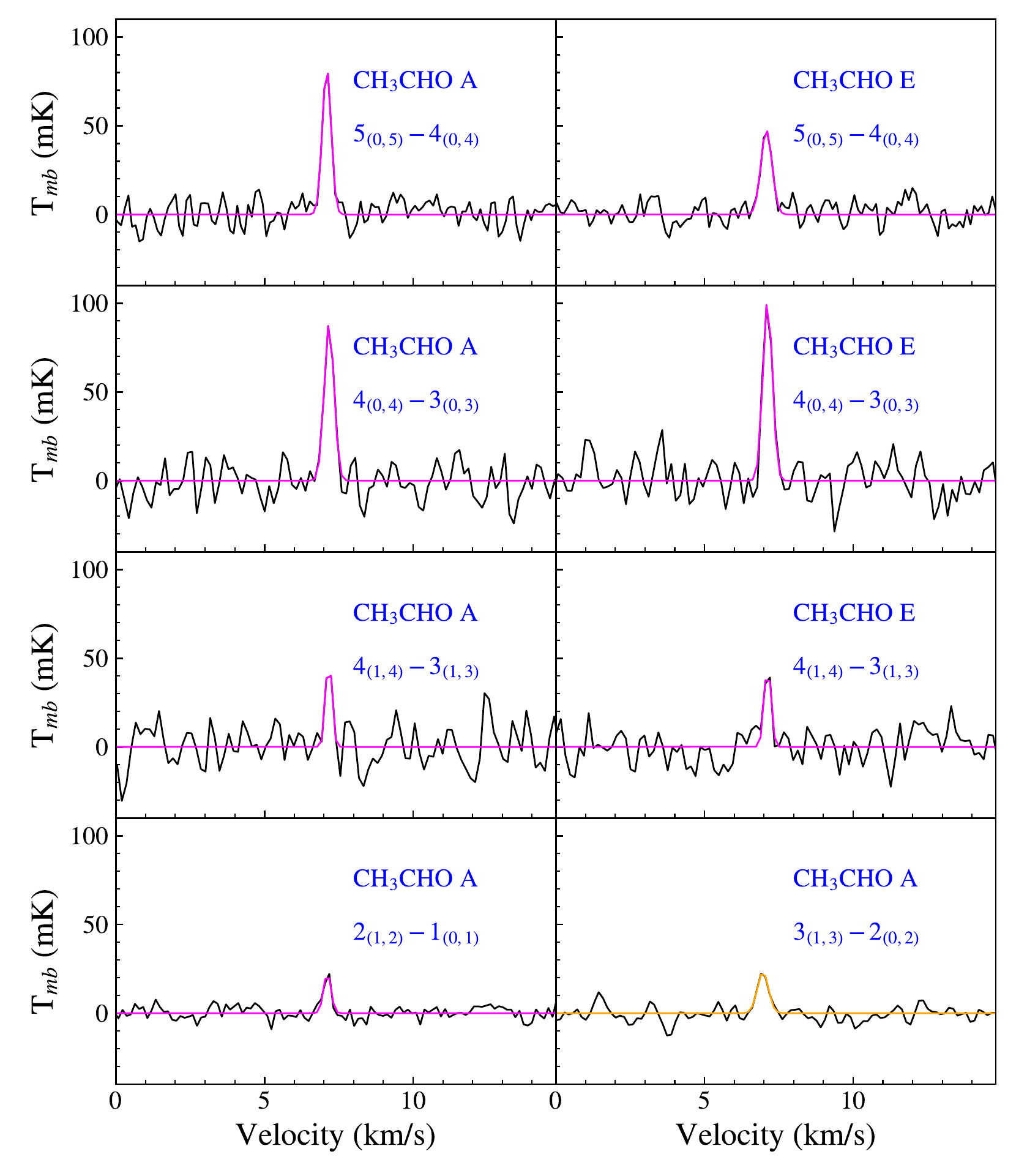} 
\end{array}$
\end{center}
\caption{\label{ch3cho_all} Spectra of all eight CH$_3$CHO transitions detected. In all panels except the bottom right the gaussian fits are overplotted in magenta, for those detected with the AROWS backend receiver. In the bottom right plot the gaussian fit is overplotted in orange and was observed with the MAC backend receiver. Each line peaks at the $\mathrm{v_{LSR}}$ ($\sim$7 km s$^{-1}$) of L1521E.}
\end{figure}

\begin{figure}
\centering
\includegraphics[width=80mm]{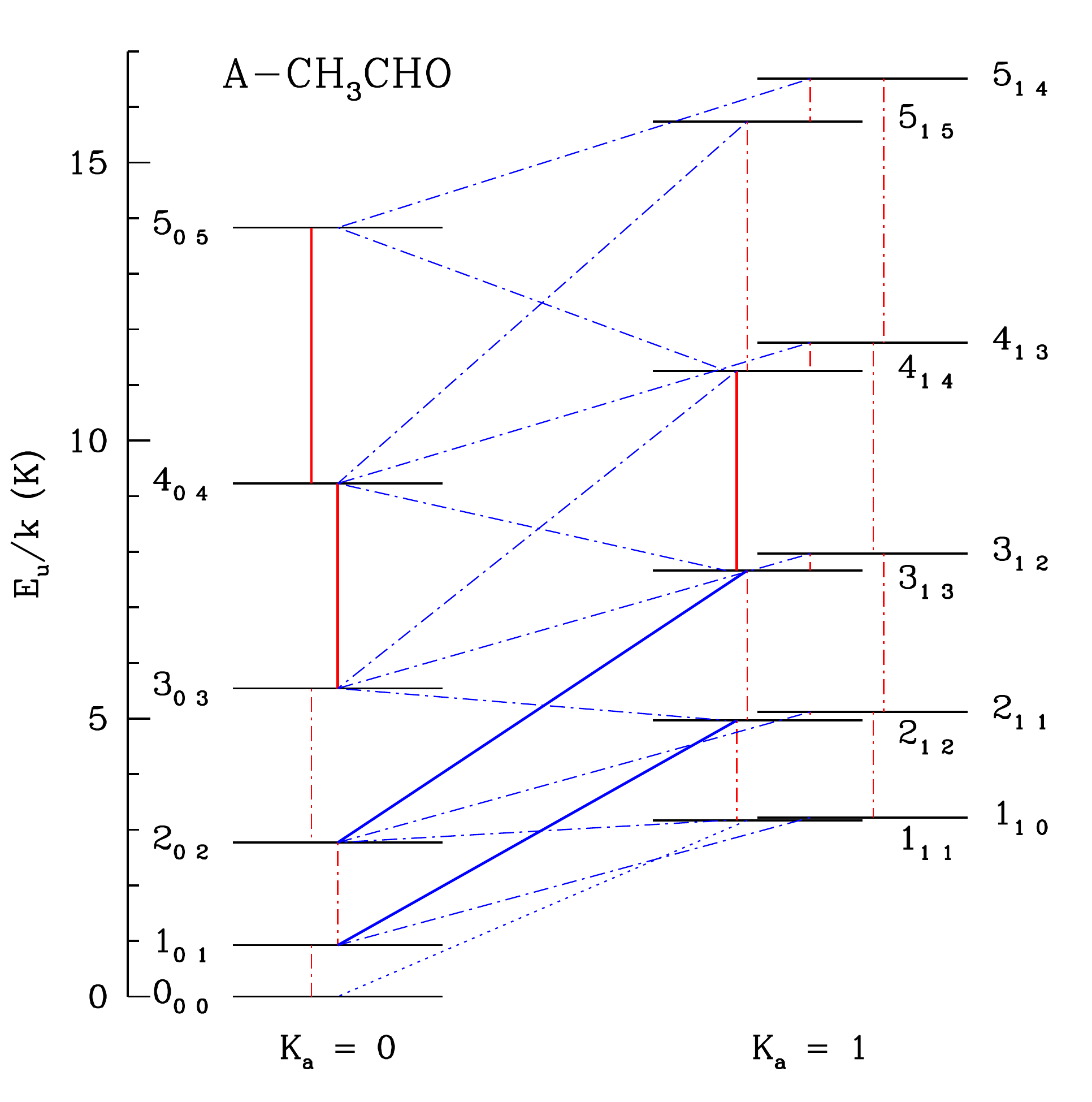} 
\caption{ \label{ACET_LVLS_fig} The rotational energy level  diagram for CH$_3$CHO A. Solid red lines connect a-type ($\Delta$K$_a$ = 0, $\Delta$K$_c$ = $\pm$1) transitions and solid blue lines connect b-type ($\Delta$K$_a$ = $\pm1$, $\Delta$K$_c$ = $\pm$1) transitions. Each solid line represents those transitions detected in L1521E. See Figure 2 in \citealt{1976ApJ...204...43G} for a more complete energy level diagram. }
\end{figure}

A total of eight transitions of CH$_3$CHO were detected by tuning to 3mm and 4mm rotational transitions suitable for 10 K gas, i.e., those with values for E$\mathrm{_u}$/k  $< 15$K (spectroscopic work detailed in \cite{1996JPCRD..25.1113K}). Observed spectra are plotted in Figure\,\ref{ch3cho_all}. With both the MAC and the AROWS backends, we observed the 96 GHz $5_{0,5}-4_{0,4}$\,A and E transitions, however we only report on the AROWS detection here. The weak CH$_3$CHO $3_{1,3}-2_{0,2}$\,A transition at 101.89\,GHz was detected with only the MAC. The remaining transitions were detected exclusively with the AROWS backend, at $\sim$0.07-0.08\,km\,s$^{-1}$ resolution. The weakest transition, CH$_3$CHO\,A\,$2_{1,2}-1_{0,1}$ at 84\,GHz, was targeted and detected with 4$\sigma$ confidence. The remaining transitions, each various K levels of J\,=\,$4 - 3$ (refer to Figure\,\ref{ACET_LVLS_fig}), were targeted with the 4mm receiver. The $76.9$\,GHz, $4_{0,4}-3_{0,3}$ transitions are brightest. Each A $\&$ E pair were observed in the same bandpass. For the $95.95$\,GHz A and E states they have a ratio of around 9:5, unlike the $76.9$ and $74.9$ GHz transitions with ratios close to the LTE statistical ratio of 1:1. Observations of A to E ratios of 1:1 have been observed
towards other sources \citep{2012A&A...541L..12B, 2017A&A...607A..20T}.

\subsubsection{Dimethyl Ether (CH$_3$OCH$_3$)}

\begin{figure}
\centering
\begin{center}$
\begin{array}{c}
\includegraphics[width=75mm]{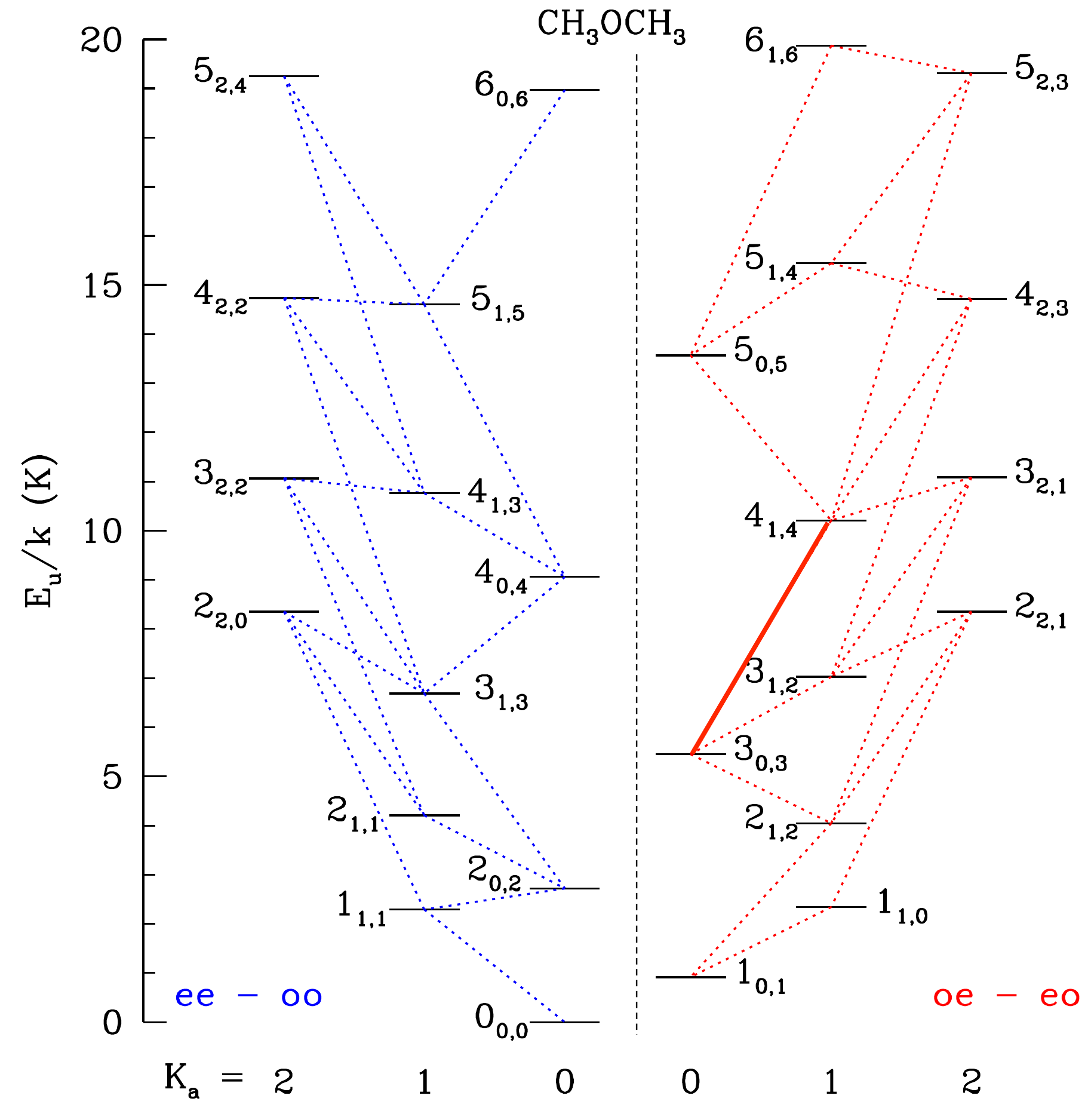} 
\end{array}$
\end{center}
\caption{ \label{DME_fig} 
The rotational energy level diagram for CH$_3$OCH$_3$ EE with only the stronger b-type transitions shown ($\Delta$K$_a$ = $\pm$1, $\Delta$K$_c$ = $\pm$1). Weaker c-type transitions are not shown ($\Delta$K$_a$ = $\pm$1, $\Delta$K$_c$ = 0). A solid red line connects the $4_{1,4}$ and $3_{0,3}$ levels, which produces the CH$_3$OCH$_3$ EE transition detected in L1521E. }
\end{figure}

A simple LTE excitation analysis of CH$_3$OCH$_3$ shows that the $4_{1,4} - 3_{0,3}$ transition at 
99.325\,GHz should be the strongest line in the 3mm band for gas at 10\,K (spectroscopic work detailed in \cite{1990ZNatA..45..702N}). Furthermore,
the upper state energy level (J$_{\rm{K_a,K_c}}$\,=\,$4_{1,4}$) can only radiate (via electric dipole radiation) through this single transition (see Figure\,\ref{DME_fig}).  
For CH$_3$OCH$_3$ there is an advantage of having torsional splitting of the rotational levels, as the lines manifest as three closely spaced lines ($\sim 2$\,MHz) with LTE statistical weights of 1:2:1 that are in a distinct, easy to identify pattern.  

\begin{figure}
\centering
\begin{center}$
\begin{array}{c}
\includegraphics[width=80mm]{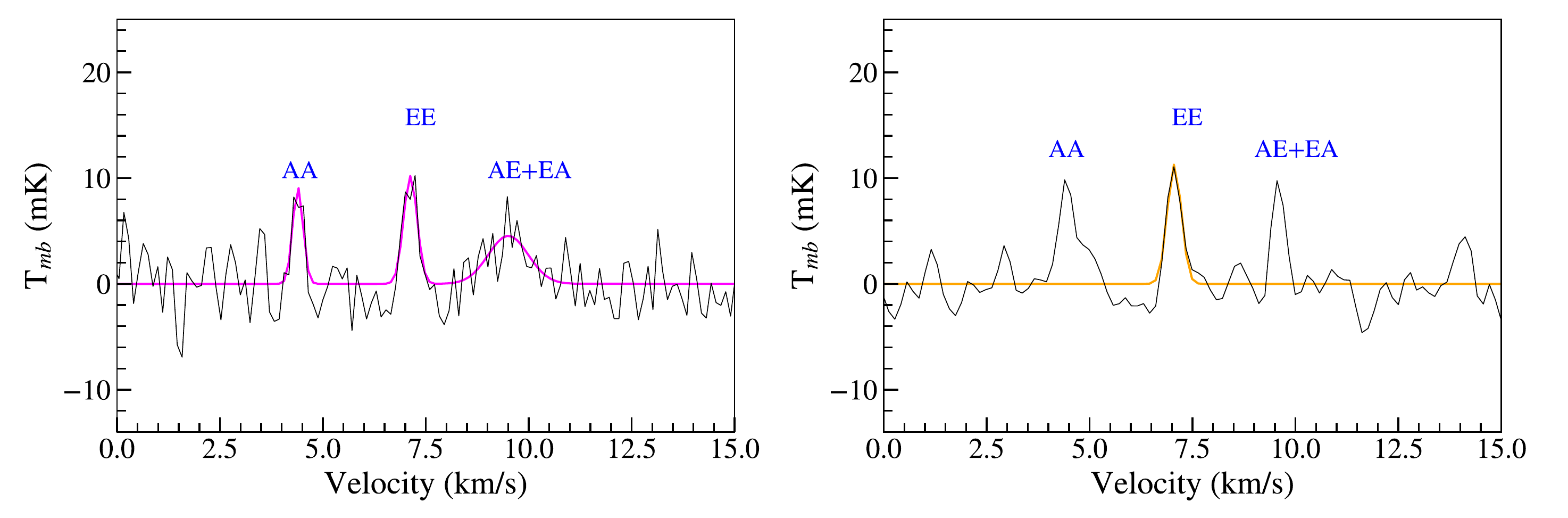}
\end{array}$
\end{center}
\caption{ \label{DME_spectra} CH$_3$OCH$_3$ observations from first observing run with the MAC backend (right) and the second observing run with the new AROWS backend (left), at a higher velocity resolution. The center EE transition peaks at the $\mathrm{v_{LSR}}$ of L1521E.}
\end{figure}

We observed CH$_3$OCH$_3$ during both observing sessions, further verifying its detection. A 93 hour (total observing time), multi-shift integration with baseline rms of 1.9\,mK using the MAC backend was successful in detecting the $4_{1,4} - 3_{0,3}$ transition (right panel of Figure\,\ref{DME_spectra}), with three distinct peaks of
signal-to-noise $> 6$ in integrated intensity at the correct frequencies for the AA, EE, and
AE+EA species of CH$_3$OCH$_3$.  During the second round of observations we integrated for 166 hours, down to an rms of 3.0 mK with the AROWS backend, confirming the previous MAC spectrometer detection (left panel of Figure\,\ref{DME_spectra}). The EE species is the brightest line detected with $\sim4\sigma$ confidence. By re-observing with AROWS we gained higher spectral resolution, resolving linewidths. Values from the AROWS observation are reported in Table \,\ref{COMresults}.

Attempts to detect the $4_{2,3} - 4_{1,4}$ and $2_{2,1} - 2_{1,2}$ transitions of CH$_3$OCH$_3$ during the first observing run were unsuccessful despite a baseline rms value of $2.2$\,mK for both lines. We calculate a $3\sigma$ upper limit to the integrated intensity of $0.8$\,mK\,km\,s$^{-1}$, assuming the
same FWHM linewidth as observed for the $4_{1,4} - 3_{0,3}$\,EE transition that was observed with the MAC backend, i.e., $\rm{FWHM} = 0.42$\,km\,s$^{-1}$. Each of the upper states of these transitions have multiple possible routes to radiate to lower energy levels. The transitions targeted in the 3mm band do not have the largest branching ratio among those transitions, which mitigate their intensity (refer to Figure\,\ref{DME_fig}).

\begin{figure}
\centering
\begin{center}$
\begin{array}{c}
\includegraphics[width=80mm]{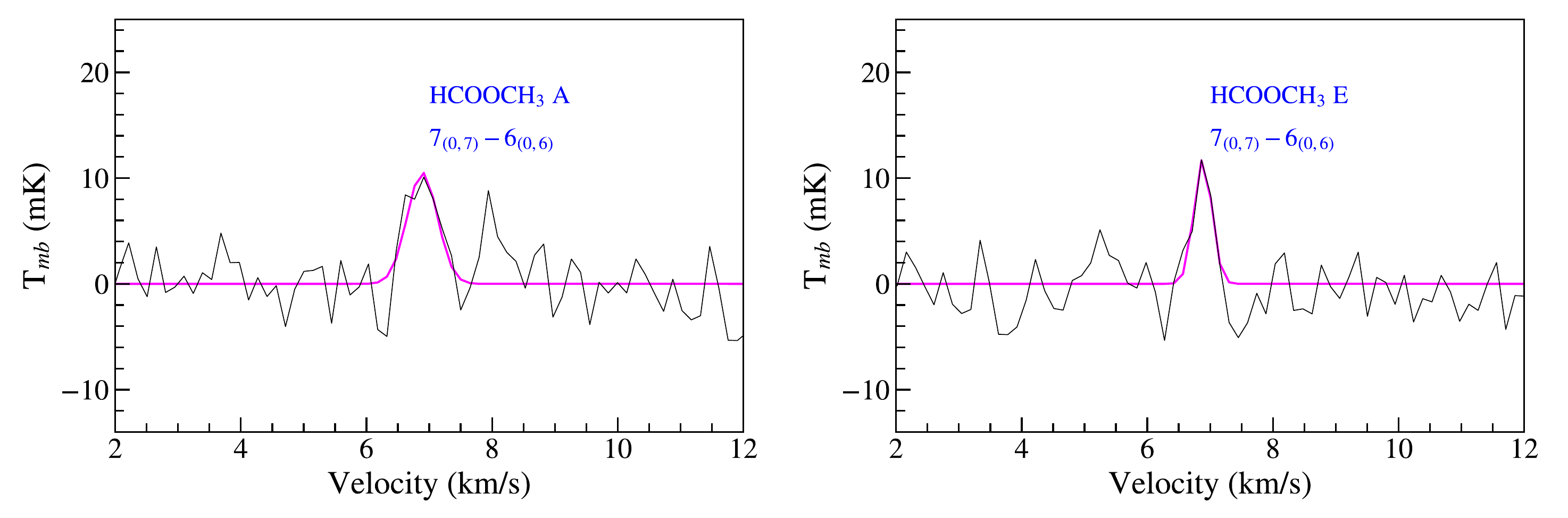}
\end{array}$
\end{center}
\caption{\label{methylform} Two transitions of the 79 GHz $7_{0,7} - 6_{0,6}$ A and E states of HCOOCH$_3$ detected in L1521E, centered on the v$_\mathrm{lsr}$ of L1521E.}
\end{figure}

\begin{figure}
\centering
\begin{center}$
\begin{array}{c}
\includegraphics[width=80mm]{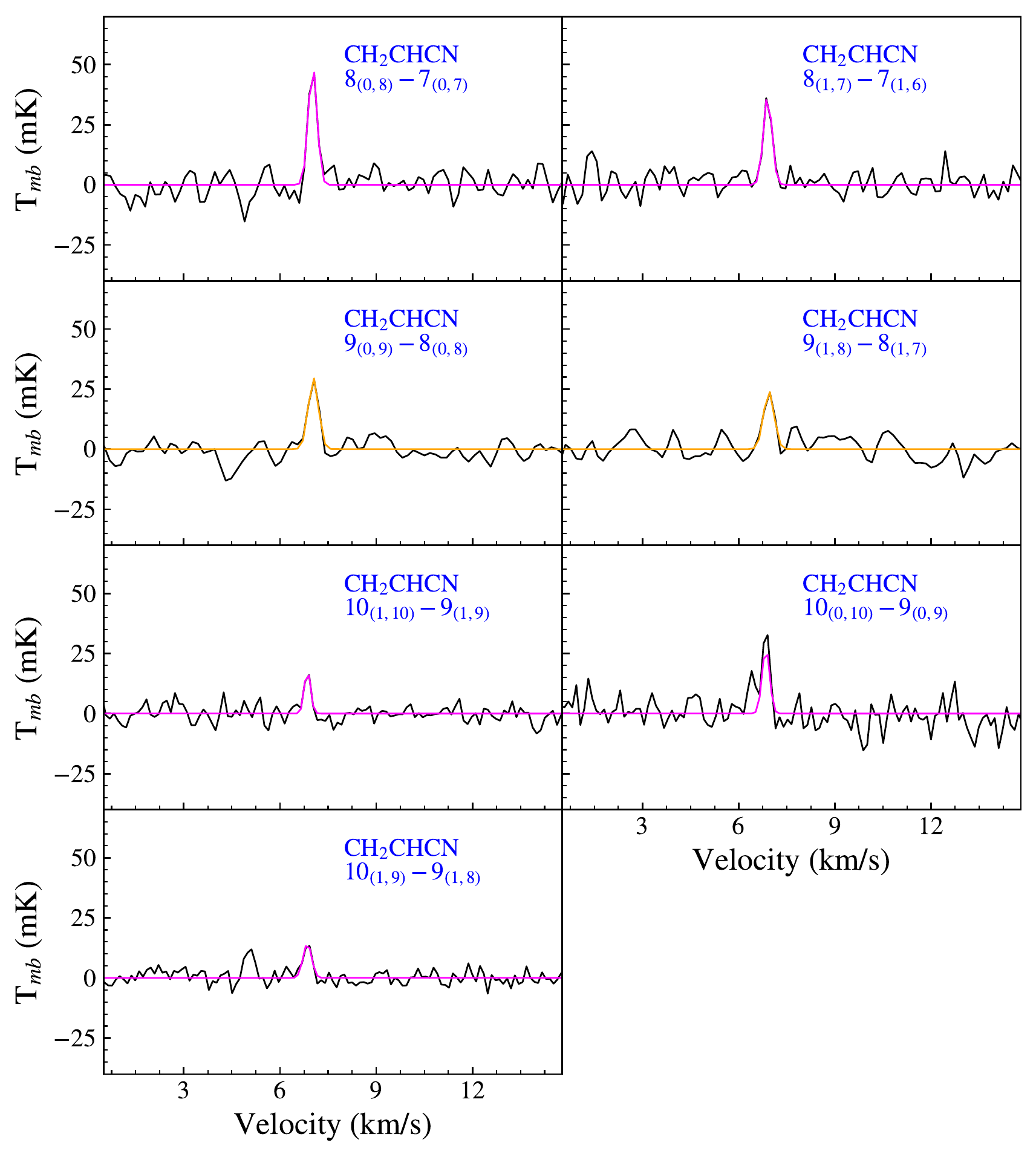} 
\end{array}$
\end{center}
\caption{\label{ch2chcn_all} Spectra of seven separate transitions of CH$_2$CHCN. We note that the J = $9-8$ transitions were detected earlier with the MAC and the gaussian fit is displayed in orange. 
Each transition is centered on the v$_\mathrm{LSR}$ of L1521E.}
\end{figure}

\begin{figure}
\centering
\begin{center}$
\begin{array}{c}
\includegraphics[width=85mm]{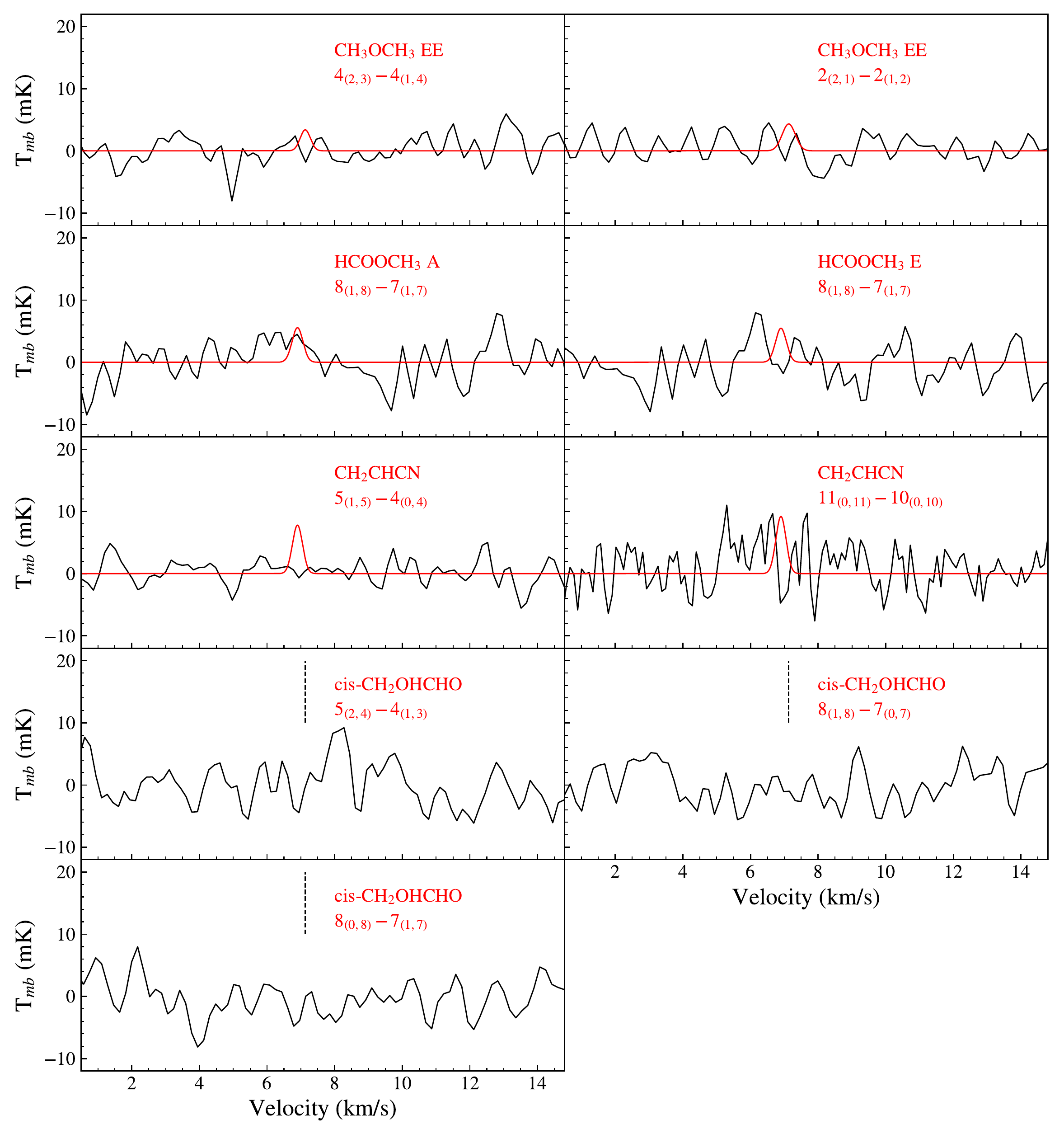} 
\end{array}$
\end{center}
\caption{\label{nondetect_spec} Spectra showing the non-detected transitions as detailed in Table\,\ref{COMresults}. For the non-detected transitions for which we have column density estimates for, their modeled Gaussian LTE spectrum is plotted in red. The attempt to detect the CH$_2$CHCN $11_{0,11}-10_{0,11}$ transition done with AROWS backend and therefore has higher velocity resolution, all others were attempted with the MAC backend. For cis-CH$_2$OHCHO, which was not detected in any transition, a vertical dashed line annotates where we would expect the line to appear, centered on the v$_\mathrm{LSR}$ of L1521E.
}
\end{figure}

\subsubsection{Methyl Formate (HCOOCH$_3$)}

Methyl formate, written as HCOOCH$_3$ or CH$_3$OCHO, was detected in the 4mm band but not in the 3mm band (spectroscopic work detailed in \cite{2009JMoSp.255...32I}).
We find the 76\,GHz transitions are more favorable for detection in cold L1521E, in part due to the lower upper energies (E$_u$/k\,=\,15.7 K), when compared to the 89 GHz transitions (E$_u$/k\,=\,20 K). We report detections of the 79 GHz $7_{0,7} - 6_{0,6}$\,A and E transitions (observed in the same bandpass), reported with $3\sigma$ confidence (Figure \,\ref{methylform}). 
The 89 GHz $8_{1,8} - 7_{1,7}$\,A and E transitions of HCOOCH$_3$ have been detected before (see \cite{2017A&A...607A..20T}), yet we only report upper limits in Table\,\ref{COMresults} after getting down to rms of 3.5\,mK. 

\subsubsection{Vinyl Cyanide (CH$_2$CHCN)} \label{subsecvnycyn}

Seven transitions of vinyl cyanide (CH$_2$CHCN) were detected and plotted in Figure\,\ref{ch2chcn_all} (spectroscopic work detailed in \cite{1959JChPh..30..777C}). Another two transitions were not detected, and we report upper limits in Table\,\ref{COMresults}. CH$_2$CHCN is the most complex organic molecule we detect with a nitrogen atom. The strongest transition detected is the $8_{0,8}-7_{0,7}$ a-type transition, with $8\sigma$ confidence. The $9_{0,9}-8_{0,8}$ transition is the second strongest, and the $9_{1,8}-8_{1,7}$ transition is slightly weaker.  

Since the upper state of the two $9 - 8$ transitions from MAC observations have similar energies ($20.4$\,K and $23.1$\,K respectively), they do not
give a strong constraint on T$\mathrm{_{ex}}$. The non-detection of the weaker $5_{1,5}-4_{0,4}$ b-type transition with a much lower upper energy state (E$_u$/k\,=\,$8.8$\,K), does give a 3$\sigma$ upper limit and constrains T$\mathrm{_{ex}} > 5 \mathrm{K}$. This is higher than the excitation temperatures we derive for CH$_3$CHO, but perhaps not
surprising since we detected even higher energy transitions, up to $10_{1,9}-9_{1,8}$ with an upper energy of E$_u$/k\,=\,27.8\,K. 

We note that CH$_2$CHCN was recently detected and mapped in Titan's atmosphere with ALMA \citep{2017AJ....154..206L}. Of astrobiological interest, CH$_2$CHCN is thought to be important as a potential molecule to form membranes in Titan's liquid methane oceans. 

\subsection{Molecular Non-Detections}\label{subsecnondetect}

Mock integrated intensity values for transitions of CH$_3$OCH$_3$, HCOOCH$_3$ and CH$_2$CHCN that were not detected were calculated based on the column density calculated from the detected transitions and assuming the same T$\mathrm{_{tex}}$ (and $f_\nu = 1$). We also plot the LTE modeled gaussian fits over the spectra in Figure\,\ref{nondetect_spec}.
For CH$_3$OCH$_3$ the $3\sigma$ upper limit to the integrated intensity is $2.4$\,mK\,km\,s$^{-1}$. The simulated integrated intensity for the $4_{2,3} - 4_{1,4}$ transition, assuming the same excitation temperature of 4.5 K, is below this limit at 1.1\,mK\,km\,s$^{-1}$. Similarly, the simulated integrated intensity for the $2_{2,1} - 2_{1,2}$ transition was 1.7\,mK\,km\,s$^{-1}$, below the $3\sigma$ upper limit.
For both $8_{1,8} - 7_{1,7}$ transitions of HCOOCH$_3$, a $3\sigma$ upper limit to the integrated intensity is calculated to be $3.9$\,mK\,km\,s$^{-1}$. The simulated integrated intensities fall below this $3\sigma$ upper limit, at 1.9 \,mK\,km\,s$^{-1}$.

In the case of CH$_2$CHCN, a $3\sigma$ upper limit to the integrated intensity is calculated to be $2.4$\,mK\,km\,s$^{-1}$ for the $5_{1,5}-4_{0,4}$ transition of CH$_2$CHCN. The simulated integrated intensity assuming the same excitation temperature of 5 K was 2.5\,mK\,km\,s$^{-1}$, which is just at our detection limit.
For the $11_{0,11}-10_{0,11}$ transition of CH$_2$CHCN a $3\sigma$ upper limit to the integrated intensity is calculated to be $1.8$\,mK\,km\,s$^{-1}$. The simulated integrated intensity assuming the same excitation temperature of 5\,K was 2.9\,mK\,km\,s$^{-1}$, which should have been detected by our observations at the 5$\sigma$ level. Yet, looking at Figure\,\ref{nondetect_spec}, we see that there is too much noise to confirm a detection. Also note the detection of this transition was attempted with the new AROWS backend and therefore has higher velocity resolution, all others were attempted with the MAC backend.

\subsubsection{Glycoaldehyde (cis-CH$_2$OHCHO)} 

We attempted to detect three separate transitions of cis-CH$_2$OHCHO but were unsuccessful. The detection limits are reported in Table \,\ref{COMresults} and spectra plotted in Figure\,\ref{nondetect_spec} (spectroscopic work detailed in \cite{1970JMoSt...5..205M} and \cite{2001ApJS..134..319B}). An especially interesting molecule to target, CH$_2$OHCHO is one of the simplest sugars \citep{1973JMoSt..16..259M}, and a constitutional isomer of both acetic acid (CH$_3$COOH) and the more ubiquitous HCOOCH$_3$.
First detected towards Sagittarius B2(N) \citep{2000ApJ...540L.107H}, as well as toward the high-mass hot core G31.41+0.31 \citep{2009ApJ...690L..93B}, CH$_2$OHCHO is likely to form mainly through surface reactions on interstellar dust grains in which HCO is an important constituent (see \cite{2015MNRAS.453.1587B, 2019MNRAS.483..806R, 2020A&A...634A..52S}). To date, cis-CH$_2$OHCHO has not been detected toward a starless core.

\begin{figure*}
\centering
\begin{center}$
\begin{array}{c}
\includegraphics[width=175mm]{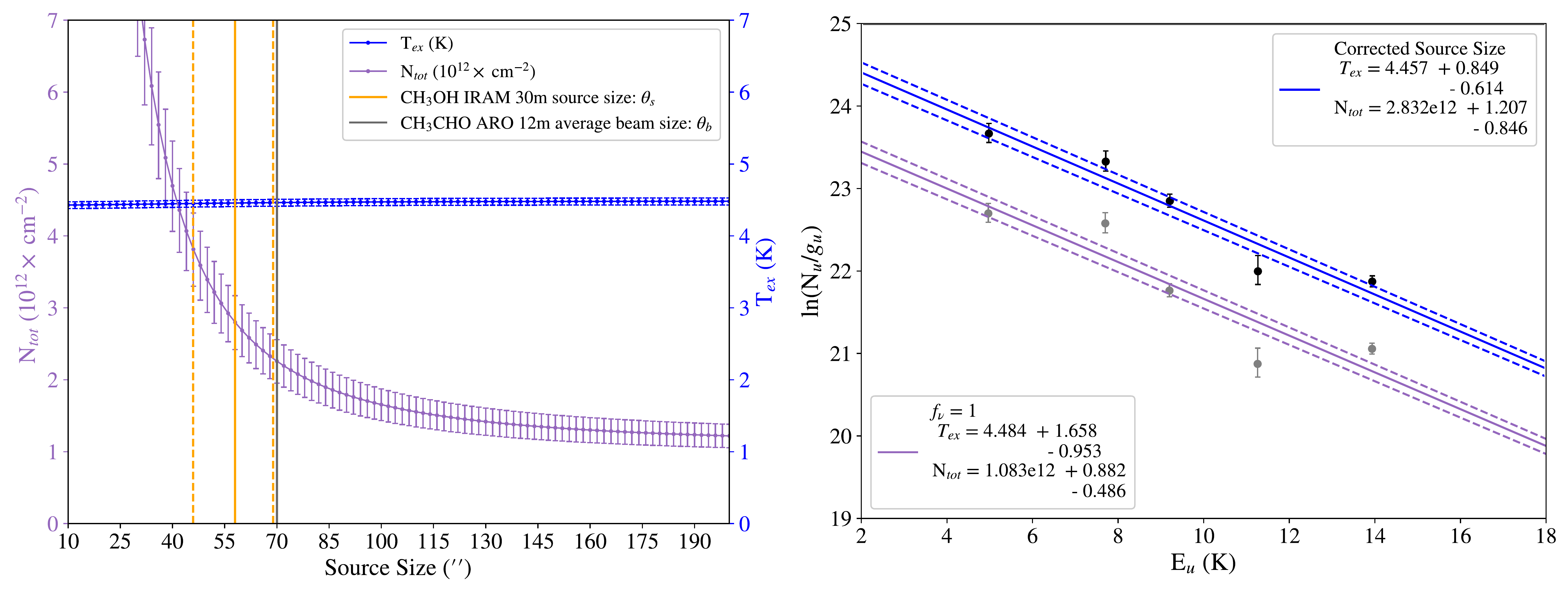} 
\end{array}$
\end{center}
\caption{ \label{fillfracplots} 
(left) The range of CH$_3$CHO A column densities as a function of source size is plotted as a purple curve with corresponding values on the left side of the y-axis. The shallow sloping blue line represents the range of CH$_3$CHO A excitation temperature as a function of source size with the corresponding values plotted on the right side of the y-axis. The orange line represents the source size calculated from a CH$_3$OH map of L1521E \citep{2019A&A...630A.136N}, and dashed lines an uncertainty window of 20$\%$ (see section \,\ref{uncert}). At this source size of $57^{''}$ (orange line), the purple curve gives the same  N$_{\mathrm{tot}}$ value as the ``Corrected Source Size" fit in the right panel. The grey line at 70$^{''}$ is the average 12m beam size for all CH$_3$CHO transitions detected. (right) Rotation diagram for CH$_3$CHO A showing the difference found if a filling fraction of 1 is assumed (grey points, purple fit) versus if the filling fraction is calculated correctly using the beam and source sizes (black points, blue fit). Instead of errors based on solely the maximum and minimum fit from observed intensity uncertainties (the dashed lines), a least squares analysis is done to find more realistic errors to our linear fit. }
\end{figure*}

\subsection{Column Density Calculations} \label{sec:colden}

For CH$_3$CHO, since the upper energy levels of the five detected A transitions span from $5.0 - 13.9$\,K, we derive meaningful limits on the excitation temperature and column density. When performing this analysis we became concerned with source size and filling fraction errors that go into these types of calculations. We include filling fraction as a parameter that goes into the column density measurement as follows, 

\begin{equation}
    \mathrm{N_u} = \frac{\rm{I}}{hA_{ul}f_\nu} \frac{u_\nu(\mathrm{T_{ex}})}{[J_\nu(\mathrm{T_{ex}}) - J_\nu(\mathrm{T_{cmb}})]} \;\;,
\end{equation}
where $f_\nu$ is the frequency (or transition) dependent filling fraction, $I$ is the integrated intensity of the line, and $u_\nu$ (Planck energy density) and $J_\nu$ (Planck function in temperature units) are defined as,
\begin{equation}
    u_\nu \equiv \frac{8 \pi h \nu^3}{c^3} \frac{1}{\exp{(h\nu/kT})-1}\;\;,
\end{equation}

\begin{equation}
    J_\nu \equiv \frac{h\nu}{k} \frac{1}{\exp{(h\nu/kT})-1}\;\;.
\end{equation}
In order to get the total column density ($\mathrm{N_{tot}}$) from the upper state value ($\mathrm{N_u}$), the Boltzmann equation in the CTEX (constant excitation temperature) approximation is used,
\begin{equation}
    \frac{\mathrm{N_u}}{g_{u}} = \frac{\mathrm{N_{tot}}}{\mathrm{Q(T_{ex}})}\exp{(-E_u/k\mathrm{T_{ex}})}\;\;,
\end{equation}
where $\mathrm{Q(T_{ex}})$ is the partition function that depends on excitation temperature, $E_u$ is the upper energy, and $g_u$ the upper state degeneracy. Rotation diagrams are constructed to get estimates for N$_\mathrm{tot}$ and T$_{\mathrm{ex}}$ (right panel of Figure\,\ref{fillfracplots}).  

Since T$_{\mathrm{ex}}$ is still a variable in these equations, an iterative fit is performed where a first a guess at T$_{\mathrm{ex}}$ is made, and then adjusted depending on the slope of the fit, until both values merge. The partition function, $Q(\mathrm{T}_{\mathrm{ex}})$, is calculated independently from lists of the rotational energy levels (see section 7 and equation 54 in  \cite{2015PASP..127..266M}).

Most calculations in the literature need to assume a filling fraction, $f_\nu$, of 1 since source size is not known a priori. The $f_\nu$ parameter is related to both the beam size, which depends on the frequency of the transition observed, and the source size, which, for our purposes, is assumed to be the same for each transition for each oxygen bearing COM molecule. Assuming that the source intensity profile ($I(\theta)$) and the telescope main beam pattern ($P_n(\theta)$) can both be approximated by a Gaussian distribution, then
\begin{equation}
    f_\nu = 
    \frac{\int I(\theta) P_n(\theta) d\Omega}{I(\theta = 0)\int P_n(\theta) d\Omega} = 
    \frac{\theta_s^2}{\theta_b^2 + \theta_s^2}\;\;,
\end{equation}
where $\theta_s$ is the source FWHM size and $\theta_b$ is the 12m FWHM beam size at the frequency of the transition. In the left panel of Figure\,\ref{fillfracplots} we plot how rotation curve fitted N$_{\mathrm{tot}}$ and T$_{\mathrm{ex}}$ change with source size for CH$_3$CHO. For example, at a source size of $57^{''}$ (orange line), the purple curve gives the same  N$_{\mathrm{tot}}$ value as the ``Corrected Source Size" fit in blue in the right panel of Figure\,\ref{fillfracplots}. With an average beam size for all five transitions at $\sim$ 70$^{\prime\prime}$, if our filling fraction is less than one, i.e. the source size is smaller than 70$^{\prime\prime}$, then N$_\mathrm{tot}$ will increase strongly with smaller source sizes. The excitation temperature, T$_{\mathrm{ex}}$, in contrast, is nearly independent of source size when beam sizes are kept consistent for one telescope dish diameter.

In order to estimate a source size, we use a CH$_3$OH map from \cite{2019A&A...630A.136N} observed with the IRAM 30m telescope. They do have a CH$_3$CHO map, however the lines were not detected with high enough sensitivity ($rms \sim$ 30 mK) to accurately match emission morphology. We know CH$_3$OH is one of the simplest and most abundant of the COMs and thus traces the extended COM structure of L1521E. Additionally, CH$_3$OH and CH$_3$CHO are thought to be linked chemically (see discussion in \,\cite{2020ApJ...891...73S}).
The area within the FWHM of the peak brightness in the CH$_3$OH map corresponds to a diameter of 62.9$^{\prime\prime}$ (source convolved with the IRAM 30m beam; $\mathrm{\theta_{src * IRAM}}$). When subtracted off in quadrature from the beam size of IRAM at the frequency of CH$_3$OH ($\mathrm{\theta_{IRAM}}$\,=\,25.6$^{\prime\prime}$), source size is found by,
\begin{equation}
    \mathrm{\theta_{src * IRAM}^2} = \mathrm{\theta_{s}^2 + \theta_{IRAM}^2}\;\;.
\end{equation}
The source size is estimated to be $\theta_{s}$\,=\,57$^{\prime\prime}$ (see section\,\ref{uncert} for uncertainties). When assuming $f_\nu = 1$, N$_\mathrm{tot}$ is underestimated by a factor of 2.6, compared to when $f_\nu$ is corrected given the proper source size (Figure\,\ref{fillfracplots}, right). Errors in the fit are calculated from a least square fitter. Our range of column densities for CH$_3$CHO A, reported in Table\,\ref{COMcolden}, correcting for source size, is then N$_\mathrm{tot}$\,$= 1.98 - 4.04 \times 10^{12}$\,cm$^{-2}$. If $f_\nu = 1$ is assumed we find N$_\mathrm{tot}$\,$= 0.59 - 1.96 \times 10^{12}$\,cm$^{-2}$. In the right panel of Figure\,\ref{fillfracplots} we also note that for $f_\nu = 1$ there is more scatter away from the linear fit compared to the data points plotted with the corrected source size. Without filling fraction corrections, column densities are underestimated.

For both CH$_3$OCH$_3$ and HCOOCH$_3$ the excitation temperatures calculated from our CH$_3$CHO analysis (T$_\mathrm{ex}$ = 4.46 K, 4.48 K) are used to derive total column densities. In order to estimate possible uncertainties introduced by this assumption, we calculate column densities given a lower T$_\mathrm{ex}$ value (at 4.0 K) and a higher T$_\mathrm{ex}$ value (at 5.5 K), finding a $\sim$15$\%$ increase and decrease, respectively, in column density, which is within our range of error. In the case of CH$_3$OCH$_3$, we only consider the well detected (4$\sigma$) EE state transition. Correcting for source size (assuming the same spatial distribution as CH$_3$OH), a plausible range of column densities are found,\,N$_\mathrm{tot}$\,$= 1.32 - 1.86 \times 10^{12}$\,cm$^{-2}$. This corresponds to abundances X(CH$_3$OCH$_3$/H$_2$)\,$= 0.82 - 1.16 \times 10^{-10}$ (see Table\,\ref{COMcolden}). In the case of HCOOCH$_3$, E state transitions went into the column density analysis, which gives a plausible range N$_\mathrm{tot}$\,$=  3.44 - 5.48 \times 10^{12}$\,cm$^{-2}$, for corrected source size. This corresponds to an abundance of X(HCOOCH$_3$/H$_2$)\,$= 2.14 - 3.42 \times 10^{-10}$. In Table\,\ref{COMcolden} column densities are also calculated for CH$_3$OCH$_3$ and HCOOCH$_3$ when $f_\nu = 1.0$ is assumed.

\begin{figure}
\centering
\begin{center}$
\begin{array}{c}
\includegraphics[width=85mm]{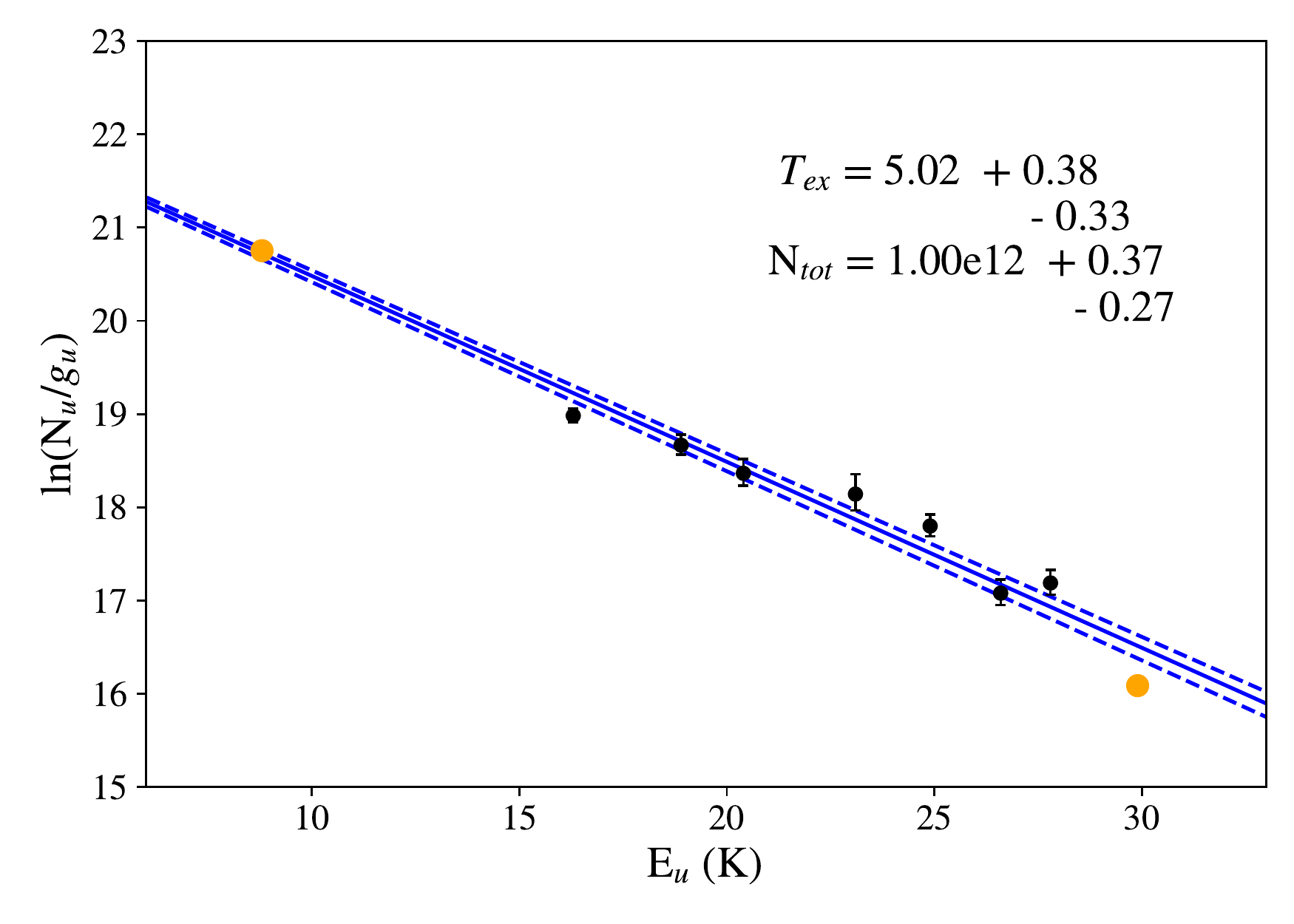} 
\end{array}$
\end{center}
\caption{ \label{vcynrotdiagram} Rotation diagram for CH$_2$CHCN constructed from seven detected transitions (in black) and two non-detections that we use as 3$\sigma$ upper limits (in orange). As in Figure\,\ref{fillfracplots}, instead of errors based on solely the maximum and minimum fit from observed intensity uncertainties (the dashed lines), a least squares analysis is done to find more realistic errors to our linear fit.}
\end{figure}

Unfortunately, for CH$_2$CHCN we are not able make the assumption that CH$_3$OH traces the same spatial structure in L1521E. Compared to oxygen bearing COMS, we would expect molecules with nitrogen to peak at different locations on the core since molecules such as CO and N$_2$H$^+$ are observed to peak differently in starless and prestellar cores (e.g., \cite{2002A&A...389L...6B, 2013A&A...560A..41L}; see also \cite{2016ApJ...830L...6J}). As a check, if we assume the same source size as for the oxygen bearing COMs we find T$_{\mathrm{ex}} < 5 \mathrm{K}$, supporting that the assumption is inaccurate. We could have used a simpler nitrogen-bearing COM, such as a CH$_3$CN map from \cite{2019A&A...630A.136N}, to estimate source size, but this map is too noisy for our purposes ($rms \sim 30 \; \mathrm{mK}$). The seven transitions of CH$_2$CHCN that were detected, as well as two upper limit constraints, are used to construct a rotation diagram in the similar manner as done for CH$_3$CHO, assuming $f_\nu = 1$ (Figure \,\ref{vcynrotdiagram}). The 3$\sigma$ integrated intensity values for the non-detection's better constrain our fit, due to the extremes in E$_u$/k at 8.8\,K and and 29.9\,K for the 89\,GHz and 103\,GHz lines, respectively. We find an excitation temperature of T$_{\mathrm{ex}} = $\,$5.02^{+0.39}_{-0.33}$\,K and range of column densities from N$_\mathrm{tot}$\,$= 0.73 - 1.37 \times 10^{12}$\,cm$^{-2}$.

\begingroup
\onecolumn 

\begin{deluxetable}{lllll}
\small
\tablecaption{Line FWHM and $v_{LSR}$ \label{COM_FWHM} }
\tablewidth{0pt}
\tablehead{
\colhead{Molecule} & \colhead{Transition} & \colhead{$\nu$} & \colhead{FHWM} & \colhead{$\mathrm{v_{LSR}}$} \\ \colhead{ } & \colhead{ } & \colhead{(GHz)} &  \colhead{km s$^{-1}$} & \colhead{km s$^{-1}$}}
\startdata
CH$_3$CHO & $3_{1,3}-2_{0,2}$ A$^{*}$ & 101.89241 &	0.44[0.07] & 6.96[0.03]\\
  & $5_{0,5}-4_{0,4}$ A & 95.96347 & 0.42[0.05] & 7.10[0.02] \\
  & $5_{0,5}-4_{0,4}$ E & 95.94744 & 0.50[0.08] & 7.08[0.03] \\
  & $2_{1,2}-1_{0,1}$ A & 84.21975 & 0.29[0.07]  & 7.10[0.02] \\
  & $4_{0,4}-3_{0,3}$ A & 76.87895 & 0.38[0.05] & 7.18[0.02] \\
 & $4_{0,4}-3_{0,3}$ E & 76.86644 & 0.37[0.04] &  7.12[0.02] \\
 & $4_{1,4}-3_{1,3}$ A & 74.89168 & 0.24[0.11] & 7.16[0.03] \\ 
 & $4_{1,4}-3_{1,3}$ E & 74.92413 & 0.27[0.10] & 7.12[0.03]   \\
CH$_3$OCH$_3$ & $4_{1,4}-3_{0,3}$ AA & 99.32607 & 0.30[0.07] & 6.97[0.03] \\
    & $4_{1,4}-3_{0,3}$ EE & 99.32522 & 0.39[0.07]  & 7.13[0.04] \\
    & $4_{1,4}-3_{0,3}$ AE+EA & 99.32436 & 1.14[0.34] & 6.93[0.14] \\
HCOOCH$_3$  &	$7_{0,7} - 6_{0,6}$ A	& 79.78389	& 0.57[0.09] & 6.96[0.05]\\
	&	$7_{0,7} - 6_{0,6}$ E	& 79.78171	& 0.32[0.08] & 6.89[0.04] 	\\
CH$_2$CHCN &	$10_{1,9} - 9_{1,8}$&	96.98245 & 0.31[0.06] & 6.85[0.03]    \\
&	$10_{0,10} - 9_{0,9}$&	94.27664	& 0.27[0.06]  & 6.85[0.02] 	\\
&	$10_{1,10} - 9_{1,9}$&	92.42626	& 0.22[0.06] &  6.86[0.02] \\
&	$9_{1,8} - 8_{1,7}$	$^{*}$&	87.31281	& 0.38[0.07] & 6.26[0.03]	\\
 &	$9_{0,9} - 8_{0,8}$ $^{*}$	&	84.94600	& 0.37[0.05] & 6.55[0.02]	\\
 & $8_{1,7}-7_{1,6}$ & 77.63384 & 0.31[0.05] & 6.91[0.02] \\
 & $8_{0,8}-7_{0,7}$ & 75.58569 & 0.32[0.04] & 7.02[0.02]  \\
\enddata
\tablecomments{ $^{*}$Transitions targeted with the MAC backend.}
\end{deluxetable}

\begin{deluxetable}{llllll}
\tablecaption{Column Densities and Abundances \label{COMcolden} }
\tablewidth{0pt}
\tablehead{
\colhead{Species:} &  \colhead{CH$_3$CHO A} & \colhead{CH$_3$OCH$_3$ EE} & \colhead{HCOOCH$_3$ E}& \colhead{CH$_2$CHCN } & \colhead{cis-CH$_2$OHCHO}}
\startdata
Corrected Source Size\\
\hline 
T$_\mathrm{ex}$  (K)   &  4.46$^{+0.85}_{-0.61}$ & 4.46  & 4.46   & \nodata& \nodata\\
N$_\mathrm{tot}$ (10$^{12}$ cm$^{-2}$)   & 2.83$^{+1.21}_{-0.85}$ & 1.59$^{+0.27}_{-0.27}$&  4.46$^{+1.02}_{-1.02}$& \nodata & \nodata\\
$^{*}$X wrt. H$_2$ (10$^{-10}$) & 1.77$^{+0.76}_{-0.53}$ & 0.99$^{+0.17}_{-0.17}$& 2.78$^{+0.64}_{-0.64}$& \nodata & \nodata \\
\hline 
$f_\nu = 1$\\
\hline 
T$_\mathrm{ex}$  (K)   & 4.48$^{+1.66}_{-0.95}$   & 4.48    &  4.48     & 5.02$^{+0.38}_{-0.33}$& 4.5 \\
N$_\mathrm{tot}$ (10$^{12}$ cm$^{-2}$)   &  1.08$^{+0.88}_{-0.49}$ & 0.74$^{+0.13}_{-0.13}$ &   1.60$^{+0.37}_{-0.37}$&  1.00$^{+0.37}_{-0.27}$&  $<$ 1.84\\
$^{*}$X wrt. H$_2$ (10$^{-10}$) &  0.68$^{+0.25}_{-0.18}$ & 0.46$^{+0.08}_{-0.08}$ & 1.00$^{+0.23}_{-0.23}$  & 0.62$^{+0.55}_{-0.31}$ &  $<$ 1.15 \\
\enddata
\tablecomments{ $^{*}$Calculated using the average H$_2$ column density of $\mathrm{N_{H_2}}$ = 1.6 $\times$ 10$^{22}$ cm$^{-2}$. Note: The excitation temperature, T$_\mathrm{ex}$, derived for CH$_3$CHO is assumed for all other oxygen-bearing species.
}
\end{deluxetable}

\endgroup 
\twocolumn

\subsection{Analysis of Uncertainties} \label{uncert}

\begin{figure}
\centering
\begin{center}$
\begin{array}{c}
\includegraphics[width=85mm]{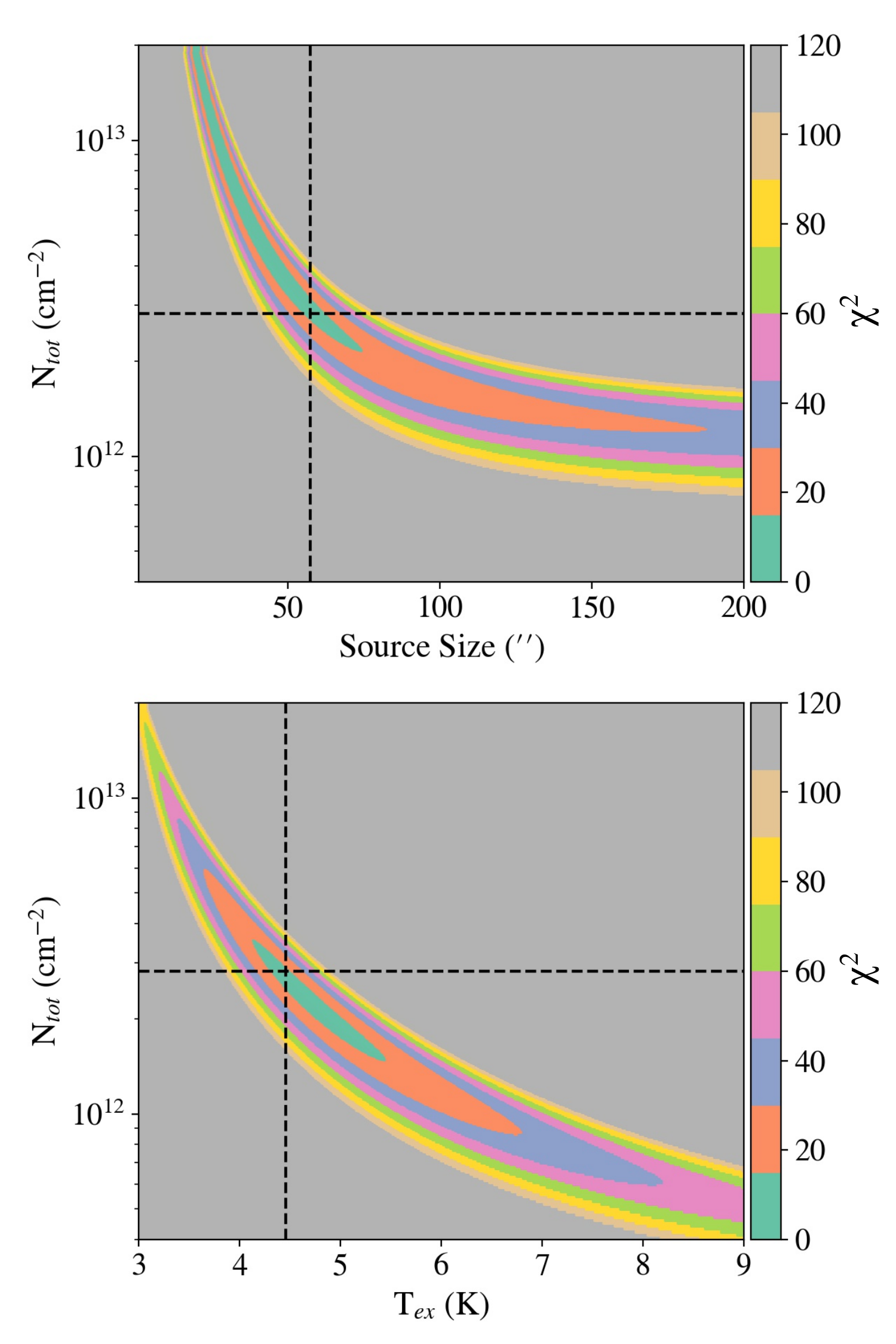}
\end{array}$
\end{center}
\caption{ \label{chisquare} The $\chi^2$-value plots of CH$_3$CHO A column density versus (top) source size (${''}$) and (bottom) excitation temperature (K). The black dashed lines represent the calculated values found from our rotation diagram analysis, given a source size of 57$^{\prime\prime}$, which crosses the minimized portion of the curve where $\chi^2 < 15$.
}
\end{figure}

We produce $\chi^2$ parameter-space plots for the determination of column density and excitation temperature from CH$_3$CHO A transitions. By re-arranging equations 2 and 5 we estimate how well the calculated line intensities fit with the observed data for each set of parameters ($\theta_{s}$, $\mathrm{T_{ex}}$, $\mathrm{N}$) by calculating, 
\begin{equation}
\chi^2 = \sum_{i=1}^{n}   (\frac{I_i^\mathrm{obs} - I_i^\mathrm{calc}}{\sigma_i^\mathrm{obs}})^2\;\;,
\end{equation}
where $n$ is the number of transitions observed, which is 5 in our case, $I_i^\mathrm{obs}$ is the observed integrated intensity, $I_i^\mathrm{calc}$ is the integrated line intensity predicted by the equations, and $\sigma_i^\mathrm{obs}$ is the uncertainty associated with the observed line integrated intensity. In the top panel of Figure \,\ref{chisquare} we plot $\chi^2$ as a function of the CH$_3$CHO A column density and source size (T$_{\mathrm{ex}}$ set to $4.46$\,K), and similarly in the bottom panel $\chi^2$ as a function of the CH$_3$CHO A column density and T$_{\mathrm{ex}}$ (source size set to $57^{''}$). In general, as column density increases, better fits (lower $\chi^2$ values) can be found for small source size and low excitation temperatures. The black dashed lines in  Figure \,\ref{chisquare} represent the calculated values from our rotation diagram analysis, which cross the minimized portion of the curve where $\chi^2 < 15$. Within this region ($\chi^2 < 15$), excitation temperature is constrained to be between $4.2-5.5\,\mathrm{K}$ and N$_{\mathrm{tot}}$ between $1.5 - 3.5\,\times\,10^{12}$\,cm$^{-2}$, comparable to the uncertainty range from our rotation diagram analysis (blue fit in right panel of Figure\,\ref{fillfracplots}). Source size, however, is not as well constrained since it is minimized asymptotically with increasing column densities given source sizes $15{''} < \theta_s <74{''}$. 

Since our $\chi^2$ analysis does not provide a strong constraint on the source size, we must adopt our original assumption that the CH$_3$OH and CH$_3$CHO source sizes are the same. In the shocks of L1157 the similar spatial distribution ($\sim 3^{''}$ scales) of both species does suggests CH$_3$OH and CH$_3$CHO are coming from the same gas \citep{2020A&A...635A..17C}. Similarly, in \cite{2019A&A...630A.136N} lower ($\sim 30^{''}$) resolution CH$_3$OH and CH$_3$CHO maps of L1521E show overlap. If we assume an error of $20\%$ in source size (see left panel Figure\,\ref{fillfracplots}) we find a range of column densities for CH$_3$CHO, N$_\mathrm{tot}$\,$= 2.3 - 3.8\,\times\,10^{12}$\,cm$^{-2}$, which is still within the errors determined from our least squares fit to our rotation diagram (Table\,\ref{COMcolden}).

\begin{figure*}
\centering
\begin{center}$
\begin{array}{cc}
\includegraphics[width=180mm]{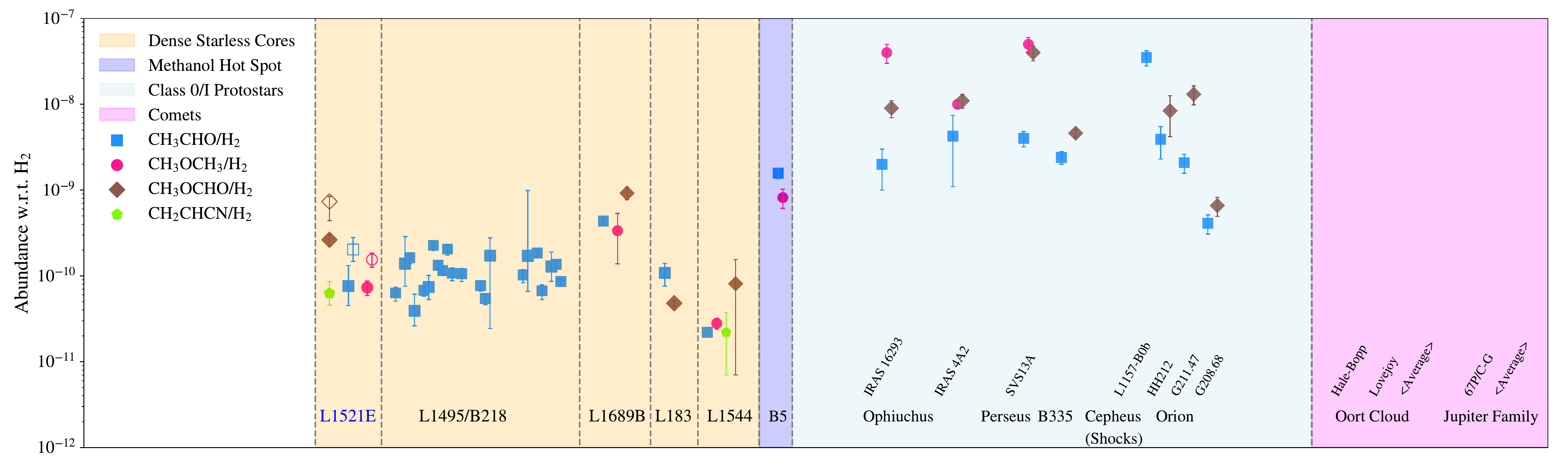}\\
\includegraphics[width=180mm]{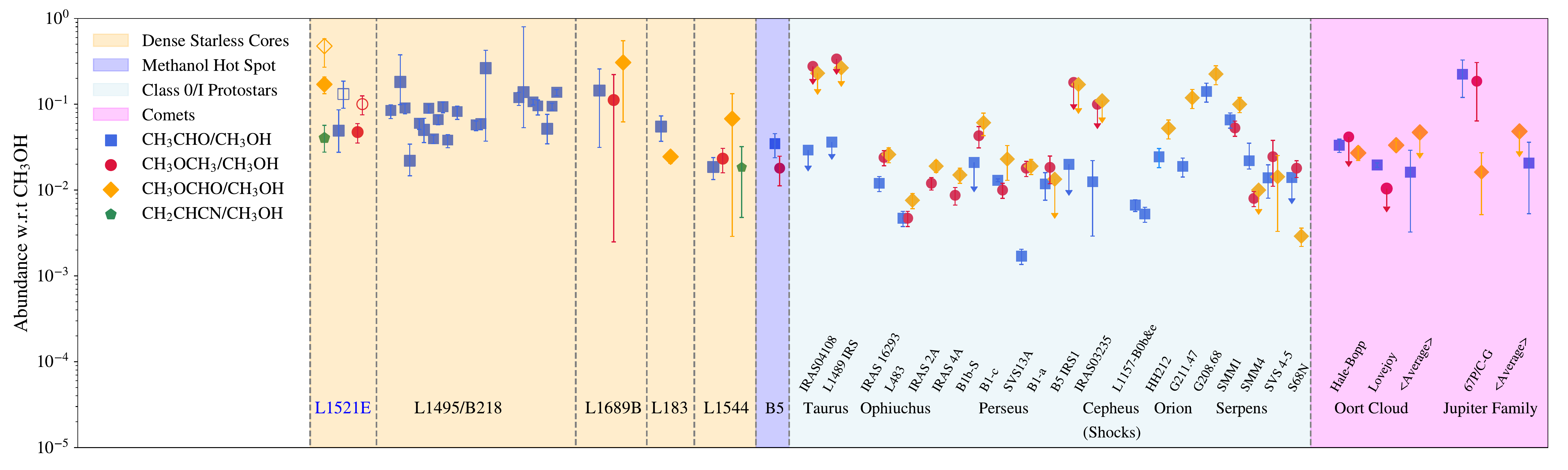}
\end{array}$
\end{center}
\caption{ \label{abundcompare} 
Comparison of COM abundances at various stages of low-mass star formation, including prestellar cores, hot cores and comets. Values for starless cores in L1495/B218 from \citealt{2020ApJ...891...73S}, L183 from \citealt{2020A&A...633A.118L}, L1689B from \citealt{2012A&A...541L..12B}, L1544 from \citealt{2016ApJ...830L...6J}, and for B5 from \citealt{2017A&A...607A..20T}. Note that for L1521E the open symbols represent values where source size was taken into account versus the filled symbols which represent values when $f_\nu = 1$ is assumed.
Class 0/I protostars arranged (roughly) by cloud they reside in and/or distance (i.e., L483 has been associated with Serpens in the past but recent GAIA data places it at more like 200-250 pc). Protostar references: B1-c, B1-bs and S68N: ALMA band 3 values from \citealt{2020A&A...639A..87V}; IRAS2A, IRAS4A: \citealt{2015ApJ...804...81T, 2017A&A...606A.121L};
IRAS 16293-2422: \citealt{2014ApJ...791...29J, 2018A&A...620A.170J}; SVS313A: \citealt{2019MNRAS.483.1850B}; IRAS03245, B1-a, B5 IRS 1, SVS 4-5, IRAS 04108, L1489 IRAS: \citealt{2016ApJ...819..140G, 2017ApJ...841..120B}; SMM1, SMM4: \citealt{2011ApJ...740...14O, 2015ApJ...804...81T}; L483: \citealt{2019A&A...629A..29J}; L1157-B0e \& L1157-B0b: \citealt{2020A&A...635A..17C}; B335: \citealt{2016ApJ...830L..37I}; HH212, G211.47, G208.68: \citealt{2019ApJ...876...63L, 2020ApJ...898..107H}.
We also include comparisons of the long-range comets, Hale-Bop and Lovejoy \citep{2019ECS.....3.1550B}, as well as Jupiter-family comet, 67P/Churyumov–Gerasimenko (67/C-G) \citep{2019ECS.....3.1854S, 2019AGUFM.P43C3489R}. Values for 67/C-G are from the mass spectroscopy instrument aboard European Space Agency's Rosetta mission, therefore isomers are indistinguishable (i.e., for CH$_3$OCHO this includes methyl formate, acetic acid and glycolaldehyde). Upper limit X(HCOOCH$_3$/CH$_3$OH) values and range of X(CH$_3$CHO/CH$_3$OH) values for all Oort Cloud Comets (OCC) and Jupiter Family Comets (JFC) also from \citealt{2019ECS.....3.1550B}.}
\end{figure*}

\section{Discussion} \label{sec:discussion}

\subsection{Abundance Comparison}\label{sec:abundcom}

We compare COM abundances to better understand how L1521E relates to other sites of low-mass star formation (Figure \,\ref{abundcompare}). Since CH$_3$OH is typically the most abundant COM, we find it interesting to investigate the abundance ratios of COMs w.r.t. CH$_3$OH, as well w.r.t. H$_2$. Overall, molecular abundances (w.r.t. H$_2$) increase between the prestellar and protostellar phase, yet the trend disappears when normalized by methanol.

The total COM abundances summed over all symmetry species are plotted in Figure\,\ref{abundcompare} for fair comparison with literature values, i.e., in the case for CH$_3$CHO a total column density of the A + E states is calculated, not just the A state as quoted in Table \,\ref{COMcolden}. Two values for each COM abundance value derived for L1521E are also plotted, one if $f_\nu = 1$ was assumed (filled symbols) and one where the source size has been corrected using the assumption of the same spatial distribution as CH$_3$OH (open symbols). 

Ordered roughly in an increasing evolutionary sequence, the molecular abundance w.r.t. H$_2$ of the starless and prestellar cores tends to increase from left to right, and then drops off after L1689B, in Figure\,\ref{abundcompare}. Most comparable to less-evolved L1521E are the `typical' starless cores in L1495/B218 as described in \cite{2020ApJ...891...73S}, which range in abundance X(CH$_3$CHO/H$_2$)\,$ = 0.4 - 2.2 \times 10^{-10}$. For prestellar core L183, embedded in the Serpens cloud, the CH$_3$CHO abundance falls in between these L1521E/L1495/B218 observations, with X(CH$_3$CHO/H$_2$)\,$ = 1.0 \times10^{-10}$ \citep{2020A&A...633A.118L}.

In the prestellar core L1689B, all COM abundances are higher than those found in L1521E, referenced from COM column density values reported in \citealt{2012A&A...541L..12B} and \citealt{2016A&A...587A.130B} assuming a $\mathrm{N(H_2)}$ value of 4$\times 10^{22}$\,cm$^{-2}$ (\citealt{2002A&A...389L...6B}). L1689B (central density $> 10^6$ cm$^{-3}$; \cite{2016A&A...593A...6S}) has been measured to have a relatively low beam-averaged CO depletion factor of N(H$_2$)$_{S_{850}}$/N(H$_2$)$_{\mathrm{C}^{17}}\mathrm{_O}$ = 2 at its center, while for well-studied L1544 the CO beam-averaged depletion factor is $>9$ at the center \citep{2003ApJ...583..789L}. L183 in Serpens (central density $> 10^{6}$ cm$^{-3}$) sits in the middle of these two cores, as it is more depleted than L1689B but not as much as L1544, with a beam-averaged CO depletion factor $\sim6$ \citep{2007A&A...467..179P}. Figure\,\ref{abundcompare} shows that the more highly evolved L1544 (central density $> 10^7$ cm$^{-3}$; \cite{2019ApJ...874...89C}) is deficient in CH$_3$CHO when compared to all other dense cores at its dust peak position with X(CH$_3$CHO/H$_2$)\,$ = 0.22\times 10^{-10}$. As speculated in \cite{2020ApJ...891...73S}, more evolved cores tend to have lower X(CH$_3$CHO/H$_2$) abundances due to the depletion of precursors to COMs and depletion of COMs in the centers of cores.

It could be, however, that environmental effects are causing the large variations between the sources, rather than evolutionary effects. L1521E, L1495/B218, and L1544 are all located within Taurus and have similar abundance values, i.e., X(CH$_3$CHO/H$_2$) varies by at most a factor of 10. As for other COMs, L1544's abundance of CH$_3$OCH$_3$ is a factor of $\sim 3$ lower than that of L1521E, with a value of X(CH$_3$OCH$_3$/H$_2$)\,$= 0.28\times 10^{-10}$. Abundances of HCOOCH$_3$ and CH$_2$CHCN are also a factor of $\sim 3$ lower, at X(HCOOCH$_3$/H$_2$)\,$= 0.81 \pm 0.74 \times 10^{-10}$ and X(CH$_2$CHCN/H$_2$)\,$= 0.22\times 10^{-10}$ \citep{2016ApJ...830L...6J}. When we compare the Taurus group (L1521E/L1495/B218/L1544) to L1689B in Ophiuchus we find, for example, X(CH$_3$CHO/H$_2$) is at most 20 times larger. Because L1689B resides in the Ophiuchus molecular cloud, it is subject to different levels of exposure from the interstellar radiation field (ISRF). The radiation field is around two times the standard ISRF for L1689B \citep{2016A&A...593A...6S}, whereas for L1544 the strength of the UV field relative to the standard ISRF has only been slightly suppressed (G$_0$ = 0.8; \citealt{2004ApJ...614..252Y}). 
Still, these ISRF intensities ($0.8-2.0$) toward the dense, obscured cores at high extinction (A$_\mathrm{V} > 10\,\mathrm{mag}$) might not be enough to affect local COM abundances. Differences in dense starless core COM abundances could be due to both their evolutionary stage as well as local environmental conditions. More observations of COMs in dense cores across many clouds are needed to say more. 

We place in between the starless core and protostar category the dense cloud B5, which has higher COM abundance (w.r.t. H$_2$) values out of all the dense cores. Known as a methanol hot spot region, B5 is not classified as prestellar or protostellar \citep{2017A&A...607A..20T}. Every starless and protostellar core plotted in Figure\,\ref{abundcompare} has a well defined dust peak in submillimeter continuum emission, yet B5 does not. Thus, we stress that it's position in between starless and protostar does not imply evolution.

The enhancement of COMs in low-mass Class 0 and Class I protostars is illustrated in Figure \,\ref{abundcompare}, and is a trend that has been seen before (e.g., \cite{2017A&A...606A.121L}). In a process laid out recently by \cite{2020A&A...641A..54C}, who observed this same trend in a large sample of massive star-forming regions, molecular abundance will increase through thermal or shock induced desorption of COMs, first formed on ices.

There is also striking similarity (within a few orders of magnitude) between COM abundances w.r.t. CH$_3$OH, which highlights similar chemical evolution between different stages of low-mass star formation. When comparing abundance values w.r.t. CH$_3$OH across all evolutionary states, we find they vary roughly within an order of magnitude. Recent observations show distributions of COM abundances similarly enhanced in the cold component of the low-mass protostars B1-c, S68N and B1-bs, indicating that COMs could be inherited from the cold prestellar core phase \citep{2020A&A...639A..87V}. While some COMs could be inherited, the ``warm-up" phase in protostars will increase COM formation efficiency \citep{2006A&A...457..927G}. 
Either CH$_3$OH is more abundant in protostellar sources or less abundant in prestellar cores when compared to other COMs (see section \,\ref{sec:formmech} for more on formation mechanisms).

Differences in individual protostar COM abundances are due to different physical conditions, since these systems are very dynamic and various heating processes from the accreting protostar, shocks, etc. can affect the local chemical environments. Figure \,\ref{abundcompare} shows that the L1157B0b shock has the highest X(CH$_3$CHO/H$_2$) abundance value  \citep{2020A&A...635A..17C}. Additionally, inhomogeneity in analysis techniques can introduce bias when comparing values in the literature. 

The recent comet measurements plotted in Figure \,\ref{abundcompare} (with individual values for Hale-Bopp and Lovejoy) include average ranges for all Oort Cloud Comets (OCC) and as well as all Jupiter Family Comets (JFC) that were reported in Table 1 \& 2 in \cite{2019ECS.....3.1550B}. One particular JFC comet, 67P/Churyumov-Gerasimenko (67P/C-G), is plotted separately and has been studied thoroughly by in situ measurements from the Rosetta mission \citep{2019ECS.....3.1854S, 2019AGUFM.P43C3489R}. Similarly to protostars, comet abundances w.r.t. CH$_3$OH are comparable (order of magnitude) as those measured in the plotted cold cores (Figure \,\ref{abundcompare}), suggesting they contain preserved material from pre-solar system material.

Overall, it is evident that similarities in abundance throughout these stages of low-mass star formation implies that the solid-state processes are already enriching the complex organic inventory
during the starless and prestellar core phase. Larger sample sets of sources (increased number of cores targeted) as well as COMs (a search for $>$ a few COMs) are needed to say much more about the link between cold phase and warm phase chemistry.

\begin{figure}
\centering
\begin{center}$
\begin{array}{c}
\includegraphics[width=82mm]{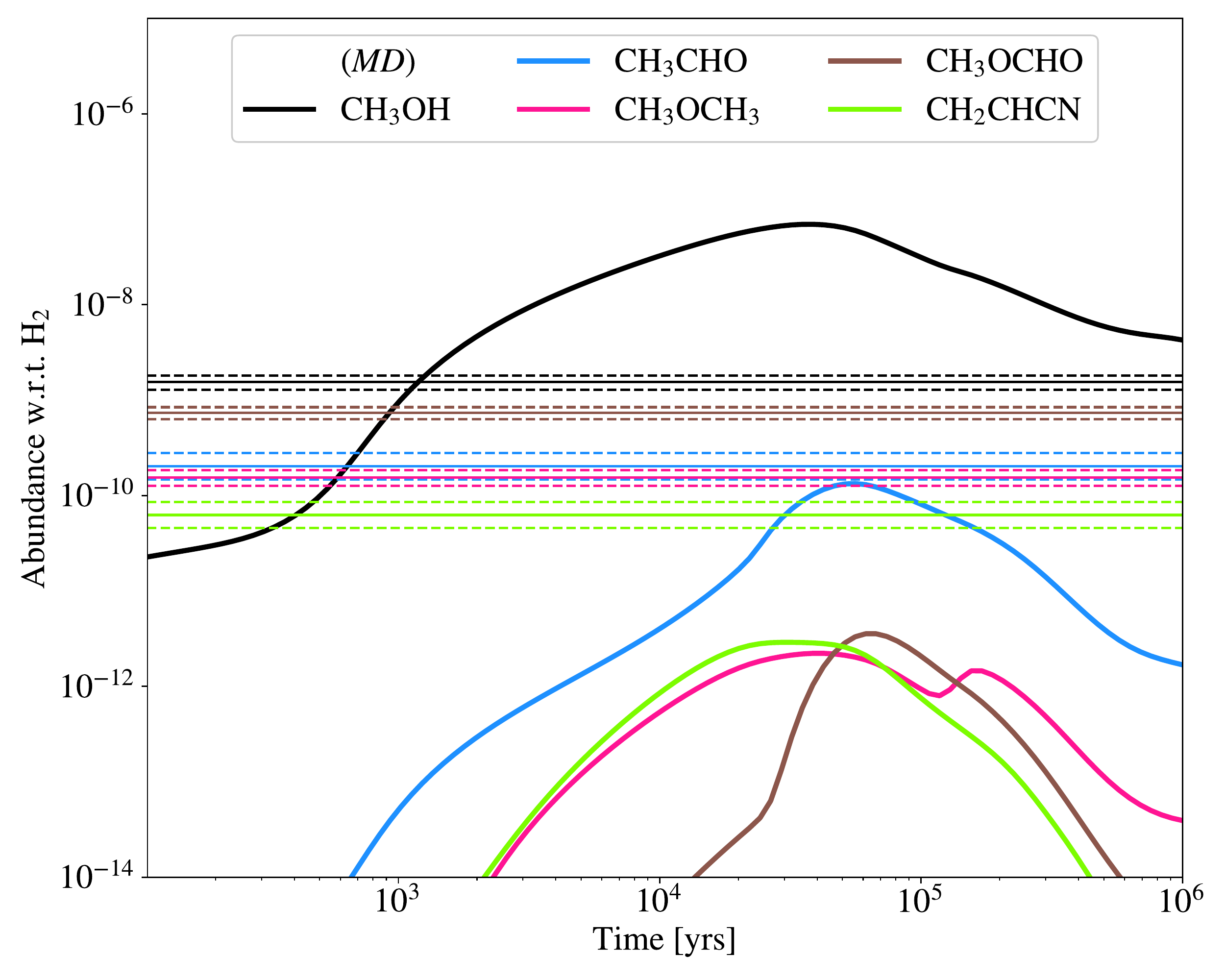} 
\end{array}$
\end{center}
\caption{ \label{models} The chemical reactive desorption model used for L1521E is described as the `Minissale$\&$Dulieu' (MD) approach, where surface reactions dictate desorption efficiency, as described in \citealt{2016A&A...585A..24M} and \citealt{2017ApJ...842...33V}. Our total (i.e., A + E states for CH$_3$CHO) calculated abundances, with uncertainties, are plotted as horizontal lines that match the color-scheme of the modeled curves. }
\end{figure}

\subsection{Formation Mechanisms} \label{sec:formmech}

Reactive desorption is thought to be one of the important mechanisms needed during the formation of COMs in cold cores. \cite{2017ApJ...842...33V} models COM distribution in well-evolved L1544 with reactive desorption of radicals followed by the rapid gas-phase formation of COMs. The results of their models show that COM
species peak in abundance on timescales of 2\,$\times 10^4 - 10^5$\,years and at radii of $0.1 - 0.3$\,pc, consistent with the position of the CO freezeout radius in L1544 (see Figures 7 and 8 of
\cite{2017ApJ...842...33V}).  At this position, the abundance of the ``surface" CO ice, defined as the top four monlayers in their models, is at its peak. 

The density of L1544 at these radii is very near the central 
density of L1521E at $3 \times 10^5$\,cm$^{-3}$.  Radiative transfer 
models of C$^{18}$O 1-0 and 2-1 are consistent with a factor of a few depletion of CO in the center of L1521E \citep{2011ApJ...728..144F}. So, if CO were depleted by a factor of two then 50\% of the gas phase CO has already frozen out of the gas phase and formed ice. A small amount of CO freeze-out onto the surface layers can drive radical formation with reactive 
desorption.
In L1521E, the CO depletion radius may essentially be unresolved by current observations and be only in the very central region of the prestellar core.
It is in these conditions, according to the models of \cite{2017ApJ...842...33V}, that the CO surface ice is maximized in abundance and COMs are formed rapidly.

We applied the chemical model described in \cite{2017ApJ...842...33V}, with several minor updates described in \cite{2019A&A...630A.136N}, to L1521E, correcting for its more shallower density profile than for L1544. A constant temperature is set to 10 K for our models presented here, and a total gas to dust mass ratio of 150:1 is assumed, to match our dust observations. In this work, we we test the `Minissale\&Dulieu' (MD) approach, which changes the reactive desorption efficiency based on ice surface composition as well as on thermodynamics and complexity of products of individual chemical reactions that occur on ice surface (detailed in \cite{2016A&A...585A..24M}). The results obtained with MD model are presented in Figure\,\ref{models}.

We find that the estimated chemical age of L1521E is indeed young, as COMs first peak $\sim 6.0\,\times$ 10$^4$ yrs, consistent with the CO depletion age of the core. Our model, however, continues to over estimate CH$_3$OH abundances by roughly two orders of magnitude, a problem seen before and described in \cite{2017ApJ...842...33V}. We find that the observed abundance for CH$_3$CHO is reproduced by this model within a factor of 2, but the abundances of CH$_3$OCH$_3$, HCOOCH$_3$ and CH$_2$CHCN are underpredicted by factors of 20-160. These COM abundances are underestimated, suggesting that we are either missing a desorption mechanism, or the current description of the already considered mechanisms should be revised in the future.

The failure of the model to fit CH$_3$OCH$_3$ and HCOOCH$_3$ may be partially due to the specific structure of L1521E since it is not as dense as other cores studied, has a low degree of CO freeze-out, and has higher sulfur abundance. Because the density is lower in comparison to L1544, the abundance of H tends to be higher than it is in dense gas. In our model, there is a reaction between atomic H and CH$_3$O to make CH$_3$ and OH. With more H, we find CH$_3$O, which is a precursor of CH$_3$OCH$_3$ and, indirectly, HCOOCH$_3$, is destroyed more efficiently leading to lower COM abundances in our model. Additionally, the model is sensitive to the elemental abundance of sulfur, affecting the abundance of atomic hydrogen on the grains through gH + gHS $\rightarrow$ gH$_2$S and gH + H$_2$S $\rightarrow$ gHS + gH$_2$ reactions. L1521E has an elevated abundance of S in order to explain observed abundances of S-bearing species in \cite{2019A&A...630A.136N}. The peak abundance of some gas-phase COMs (e.g., CH$_3$OCH$_3$) increases when the elemental abundance of S is decreased.

As for CH$_2$CHCN, it is not reproduced by our model because there are no known efficient routes of its formation in the gas phase (see e.g. KIDA database; \cite{2012ApJS..199...21W}). When CH$_2$CHCN is formed on grains it does not desorb efficiently to the gas in M$\&$D model, because this is a relatively polyatomic species, and energy released during the formation of CH$_2$CHCN is redistributed between many degrees of freedom in the molecule making the reactive desorption unlikely.

The high COM abundances in the low density starless core L1521E reveals the importance of alternative formation mechanisms, like non-diffusive chemistry. The non-diffusive chemical models, tailored for L1544, from \cite{2020ApJS..249...26J}, do show that 3-body surface reactions can enhance COMs in the gas-phase up to the levels we see in L1521E as well. In their time-dependent models, CH$_3$OCH$_3$ is under produced while HCOOCH$_3$ is over produced (see their Figure 2). However, once they optimize the three-body excited-formation efficiency (3-BEF), originally assumed to proceed with 100$\%$ efficiency, HCOOCH$_3$ can be matched to the data. Unlike in our chemical reactive desorption models, they do not have the problem of over producing CH$_3$OH.

These chemical models have many uncertainties (i.e. extrapolation of reaction rates to 10 K, etc.), and there has not been a complete study of the initial conditions or
full parameters space (see \cite{2004AstL...30..566V, 2008ApJ...672..629V, 2005A&A...444..883W, 2006A&A...451..551W, 2010A&A...517A..21W}).
Despite not completely understanding the formation of COMs, we can still propose a basic scenerio
where the core reaches densities of
$\sim 10^5$\,cm$^{-3}$ and CO freezeout rapidly increases, then COMs are formed rapidly
assisted by the reactive desorption of radicals and CH$_3$OH in the ice surface layers.  As the central density
increases, the COMs also begin to freeze out of the gas phase. The CO depletion radius moves out
in the core as does the region of peak COM formation. Thus, peak COM formation proceeds in a wave that moves outward with time following the CO depletion radius as it propagates through the 
envelope. 
When the protostar forms, then the COMs that have frozen out in the densest regions of the core are released from the ices back into the gas phase from the inside outward.

How can we further test which physical mechanism is
responsible for the formation of COMs in cold cores? 
CH$_3$OCH$_3$ is an interesting molecule to try and explain in cold environments, due to our current understanding of grain-surface radical chemistry.
As demonstrated in \cite{2016ApJ...821...46T} and \cite{2019MNRAS.482.3567S}, the rate coefficients for the formation process of CH$_3$OCH$_3$ by purely gas-phase, ion-molecule routes are not efficient in cold objects like prestellar cores. 
The release of CH$_3$OCH$_3$ from the grains to the gas is problematic as laboratory experiments show that it would rather break the molecule \citep{2016A&A...585A..24M}.
One test is to look for the radical precursors that must
be present in the gas phase to form the COMs. 

Key reactions that are required for the formation
of CH$_3$CHO in our model include the relative slow radiative association of CH$_3$ + HCO and the neutral-neutral
reaction of CH + CH$_3$OH \citep{2000PCCP....2.2549J}. HCO and CH$_3$OH have been detected previously toward L1521E and are abundant ($> 10^{-10}$) in the gas phase \citep{2016A&A...587A.130B, 2019A&A...630A.136N}. CH has its ground state, lambda doubled hyperfine transitions in its lowest energy ladder at $3.3$ GHz, but no observations toward L1521E have been published. CH$_3$ is a planar molecule with no permanent electric dipole moment and cannot be easily observed in cold gas. Recent work has dismissed the formation of CH$_3$CHO from CH$_3$OH due to newly computed rate constants, instead favoring a formation route that starts from ethanol and the OH radical \citep{2020MNRAS.499.5547V}, further motivating an investigation into alternative COM formation schemes. 

An important reaction for the formation of CH$_3$OCH$_3$ appears
to be the radiative association of CH$_3$O + CH$_3$ (i.e., \cite{2013ApJ...769...34V, 2015MNRAS.449L..16B}). The CH$_3$O radical was not detected towards L1521E, constraining column density $\mathrm{N} < 2.0 \times 10^{11}$ and X(CH$_3$O/HCO)\,$< 0.10$ \citep{2016A&A...587A.130B}. For HCOOCH$_3$, which exhibits very similar radial and temporal behavior as CH$_3$OCH$_3$ in the models (Figure\,\ref{models}), a key chemical species is the hydroxyl radical, OH, which has been shown experimentally to react with CH$_3$OH to form key radical CH$_3$O. Our models here, and historically (e.g., \cite{2013ApJ...769...34V}), have underestimated abundances of HCOOCH$_3$. However, recent studies have shown that HCOOCH$_3$ abundances can be increased by the addition of cosmic ray driven reactions \citep{2018ApJ...861...20S} and, as previously mentioned, through the non-diffusive surface chemistry \citep{2020ApJS..249...26J}. 

From comparison of Figure\,\ref{abundcompare} and Figure\,\ref{models}, one can conclude that ``viable" scenarios of formation of COMs in star-forming regions must fulfill two constraints. First, models based on such scenarios must be able to reproduce observed fractional abundances of COMs w.r.t. hydrogen. Second, gas-phase abundance ratios between COMs and CH$_3$OH must be reproduced. The universality of COMs-to-methanol gas-phase abundance ratios revealed in Figure\,\ref{abundcompare}, supported by the fact that unlike H$_2$, abundance of CH$_3$OH can be derived directly in a consistent way with abundances of COMs, suggests very high importance of the second constraint. Our model, as described by \cite{2017ApJ...842...33V}, does not fulfill these constraints for L1521E. Thus, the gas-phase scenario of cold COM formation is not sufficient in some cases. These results suggest that COMs observed in cold gas formed not only in gas-phase reactions, but also on surfaces of interstellar grains. Currently very little is known about exact mechanisms of transport of COMs possibly formed on cold grains (T\,$\sim$\,10K) to the gas phase.

The detection of a rich COM chemistry in young cold core L1521E presents an interesting challenge for future modeling efforts, requiring some type of unified approach combining cosmic ray chemistry, reactive desorption and non-diffusive surface reactions. 

\section{Conclusions} \label{sec:conclusion}

Observations of COMs in the young starless core, L1521E, sheds light on the initial chemical conditions at one of the earliest stages of star formation. From our molecular line survey we find;
\begin{enumerate}
    \item The first detections of CH$_3$OCH$_3$, HCOOCH$_3$ and CH$_2$CHCN in L1521E.
    Abundance ranges for each of these species are provided; i.e., X(CH$_3$OCH$_3$/H$_2$)\,$= 0.82 - 1.16 \times 10^{-10}$, X(HCOOCH$_3$/H$_2$)\,$= 2.14 - 3.42 \times 10^{-10}$, and X(CH$_2$CHCN/H$_2$)\,$= 0.31 - 1.17 \times 10^{-10}$.
    \item A strong constraint on column density and T$_\mathrm{ex}$ (used to calculate abundances for the other oxygen-bearing COMs) of CH$_3$CHO from a rotation diagram analysis of five A-state transitions, gives an abundance range X(CH$_3$CHO/H$_2$)\,$= 1.24 - 2.53 \times 10^{-10}$.
    \item Without taking into account source size when calculating column densities (i.e., assuming a filling fraction of 1), factors of a few in uncertainty arise. Thus, we stress the importance of high resolution maps of COMs in upcoming surveys.
    \item The connection between L1521E, other starless and prestellar cores, low-mass hot cores and comets suggest that some COMs are seeded early on in the star formation process.
    \item The modeled COM formation mechanisms suggest that radical desorption plus gas-phase chemistry alone is not enough to account for the formation of COMs in the cold, dense environments native to prestellar cores.
    \item We suggest that the ratio of gas-phase abundances of COMs to the abundance of methanol is an important constraint for scenarios of formation of complex organic species in star-forming regions.
\end{enumerate}

\section*{Acknowledgments}
We thank our referee, Victor Rivilla, for their insightful comments and for improving the quality of this manuscript. We also thank Paola Caselli, Claudio Codella, Miwha Jin and Rob Garrod for their valuable conversations and insights that improved this paper.
We are thankful that we have the opportunity to conduct astronomical research on Iolkam Du'ag (Kitt Peak) in Arizona and we recognize and acknowledge the very significant cultural role and reverence that these sites have to the Tohono O'odham Nation.
We sincerely thank the operators of the Arizona Radio Observatory (Michael Begam, Kevin Bays, Clayton Kyle, and Robert Thompson) for their assistance with the observations. 
The 12 m Telescope is operated by the Arizona Radio Observatory (ARO), Steward Observatory, University of Arizona, with funding from the State of Arizona, NSF MRI Grant AST-1531366 (PI Ziurys), NSF MSIP grant SV5-85009/AST- 1440254 (PI Marrone), NSF CAREER grant AST-1653228 (PI Marrone), and a PIRE grant OISE-1743747 (PI Psaltis).
Yancy Shirley and Samantha Scibelli were partially supported by NSF Grant AST-1410190 (PI Shirley). 
Samantha Scibelli is supported by
National Science Foundation Graduate Research Fellowship (NSF GRF) Grant DGE-1143953. Anton Vasyunin is supported by the Russian Ministry of Science and Higher Education via the State Assignment Project FEUZ-2020-0038.

\section*{Software}

\href{http://dx.doi.org/10.1051/0004-6361/201322068}{Astropy} (\cite{2013A&A...558A..33A}, \cite{2018AJ....156..123A}), \href{DOI:10.1109/MCSE.2011.37}{NumPy} \citep{2020arXiv200610256H}, \href{http://www.scipy.org}{SciPy} \citep{2020SciPy_NMeth}, \href{http://dx.doi.org/10.1109/MCSE.2007.55}{Matplotlib} \citep{2007CSE.....9...90H}, \href{https://www.iram.fr/IRAMFR/GILDAS/}{GILDAS CLASS} (\cite{2005sf2a.conf..721P}, \cite{2013ascl.soft05010G})

\section*{Data availability}

The data underlying this article will be shared on reasonable request to the corresponding author.


\bibliographystyle{mnras}
\bibliography{COM_L1521E_mnras}




\bsp	
\label{lastpage}
\end{document}